\documentclass[twocolumn,showpacs,prb,aps]{revtex4}
\usepackage{amsmath}
\usepackage{graphicx}
\usepackage{color}
\newcommand{\be}{\begin{equation}}
\newcommand{\ee}{\end{equation}}
\newcommand{\bea}{\begin{eqnarray}}
\newcommand{\eea}{\end{eqnarray}}
\newcommand{\bs}{\begin{split}}
\newcommand{\bse}{\begin{subequations}}
\newcommand{\ese}{\end{subequations}}
\newcommand{\cacoas}{${\rm CaCo_{1.86}As_2}$}
\newcommand{\cacoastwo}{${\rm CaCo_2As_2}$}

\begin{document}
\title{Physical Properties of Metallic Antiferromagnetic CaCo$_{\bf 1.86}$As$_{\bf 2}$ Single Crystals}
\author{V. K. Anand}
\altaffiliation{Present address: Helmholtz-Zentrum Berlin f\"ur Materialien und Energie, Hahn-Meitner Platz~1, 14109 Berlin, \mbox{Germany}.}
\author{R. S. Dhaka}
\altaffiliation{Present address: Department of Physics, Indian Institute of Technology Delhi, Hauz Khas,
New Delhi-110016, India}
\author{Y. Lee}
\author{B. N. Harmon}
\author{Adam Kaminski}
\author{D. C. Johnston}
\altaffiliation{johnston@ameslab.gov}
\affiliation{Ames Laboratory and Department of Physics and Astronomy, Iowa State University, Ames, Iowa 50011, USA}

\date{\today}

\begin{abstract}

X-ray powder diffraction (XRD), magnetic susceptibility $\chi$, isothermal magnetization $M$, heat capacity $C_{\rm p}$ and in-plane electrical resistivity $\rho$ measurements as a function of temperature $T$ and magnetic field $H$ are presented for \cacoas\ single crystals.  The electronic structure is probed by angle-resolved photoemission spectroscopy (ARPES) measurements of \cacoas\ and by full-potential linearized augmented-plane-wave calculations for the supercell ${\rm Ca_8Co_{15}As_{16}}$ (${\rm CaCo_{1.88}As_2}$).   Our XRD crystal structure refinement is consistent with the previous combined refinement of x-ray and neutron powder diffraction data showing a collapsed-tetragonal ${\rm ThCr_2Si_2}$-type structure with 7(1)\% vacancies on the Co sites corresponding to the composition \cacoas\ [D.~G.~Quirinale et~al., Phys.~Rev.~B {\bf 88}, 174420 (2013)].  The anisotropic $\chi(T)$ data are consistent with the magnetic neutron diffraction data of Quirianale~et~al.\ that demonstrate the presence of A-type collinear antiferromagnetic order below the N\'eel temperature $T_{\rm N} = 52$(1)~K with the easy axis being the tetragonal $c$~axis.  However, no clear evidence from the $\rho(T)$ and $C_{\rm p}(T)$ data for a magnetic transition at $T_{\rm N}$ is observed.  A metallic ground state is demonstrated from the band calculations and the $\rho(T)$, $C_{\rm p}(T)$ and ARPES data, and spin-polarized calculations indicate a competition between the A-type AFM and FM ground states.  The $C_{\rm p}(T)$ data exhibit a large Sommerfield electronic coefficient reflecting a large density of states at the Fermi energy ${\cal D}(E_{\rm F})$ that is enhanced compared with the band structure calculation where the bare ${\cal D}(E_{\rm F})$ arises from Co $3d$ bands.  At 1.8~K the $M(H)$ data for $H \parallel c$ exhibit a well-defined first-order spin-flop transition at an applied field of 3.5~T\@.  The small ordered moment of $\approx 0.3~\mu_{\rm B}$/Co obtained from the $M(H)$ data at low~$T$, the large exchange enhancement of $\chi$ and the lack of a self-consistent interpretation of the $\chi(T)$ and $M(H,T)$ data in terms of a local moment Heisenberg model together indicate that the magnetism of \cacoas\ is itinerant.

\end{abstract}


\maketitle

\section{\label{Intro} INTRODUCTION}

The discovery of high-temperature superconductivity in iron arsenides that share features with the high-$T_c$ cuprate superconductors such as antiferromagnetic (AFM) ordering in the nonsuperconducting parent compounds and a transition metal square sublattice has attracted significant attention in seeking a common understanding of the mechanism of high-$T_c$ in iron arsenides and cuprates.\cite{Rotter2008a, Chen2008a, Sasmal2008, Wu2008, Sefat2008, Torikachvili2008, Ishida2009, Alireza2009, Johnston2010, Canfield2010, Mandrus2010, Johnston1997, Damascelli2003, Lee2006} On the other hand, in contrast to the parent cuprate compounds which are local-moment AFM insulators, the parent iron arsenide compounds such as $A$Fe$_2$As$_2$ ($A$ = Ca, Sr, Ba) have an AFM ground state in which an itinerant spin-density wave transition at a N\'eel temperature $T_{\rm N}$ is closely accompanied by a structural distortion from the body-centered tetragonal ${\rm ThCr_2Si_2}$-type (122-type) structure to an orthorhombic structure.  The ordered moment in the collinear ``stripe'' AFM structure is oriented in the tetragonal $ab$ plane,\cite{Johnston2010} breaking the fourfold tetragonal rotational symmetry about the $c$ axis, which is correlated with the orthorhombic lattice distortion.  Superconductivity in the $A$Fe$_2$As$_2$ materials occurs upon suppression of the SDW transition by partial substitutions at the $A$, Fe and/or As-sites or by the application of external pressure.\cite{Johnston2010}  Superconductivity also occurs in 122-type compounds not containing Fe such as found from recent measurements of ${\rm CaPd_2As_2}$ ($T_{\rm c} = 1.27$~K) and ${\rm SrPd_2As_2}$ ($T_{\rm c} = 0.92$~K) single crystals.\cite{Anand2013}

In our effort to search for and understand the physics of novel 122-type arsenide compounds, we recently investigated the physical properties of ${\rm SrCu_2As_2}$ and ${\rm CaCu_{1.7}As_2}$ which form in a collapsed tetragonal (cT) structure and behave like $sp$-band metals with Cu atoms in nonmagnetic $3d^{10}$ Cu$^{+1}$ electronic configuration with the atomic $4s$~valence electron being itinerant.\cite{Anand2012a,Anand2012b} In the cT structure ${\rm SrCu_2As_2}$ has a $c/a$ ratio of 2.39 in contrast to the $c/a$ ratio of 3.29 for the uncollapsed tetragonal (ucT) structure of ${\rm BaFe_2As_2}$.\cite{Anand2012a}  The associated dimerization of the As atoms of adjacent layers of the cT structure leads to a formal As$^{-2} \equiv $ [As--As]$^{-4}$/2 oxidation state in contrast to the oxidation state As$^{-3}$ in ${\rm BaFe_2As_2}$. Thus with Sr in the $+2$ oxidation state, the formal oxidation state of Cu in ${\rm SrCu_2As_2}$ is Cu$^{+1}$ with a nonmagnetic $3d^{10}$ electronic configuration, leading to the observed $sp$-band metal behavior\cite{Anand2012a} as predicted from electronic structure calculations.\cite{Singh2009}

A pressure-induced lattice collapse transition from the ucT to the cT phase has been found to quench the magnetic ordering of Fe magnetic moment and associated AFM fluctuations in ${\rm CaFe_2As_2}$.\cite{Kreyssig2008, Goldman2009, Pratt2009} On the other hand, Co$^{+1}$ with a $3d^{8}$ electronic configuration in cT phases is magnetic.\cite{Reehuis1998, Jia2009, Jeitschko1985, Reehuis1990, Morsen1988, Reehuis1993, Reehuis1994, Huhnt1997, Chefki1998}  Thus the behaviors of the Co- and Fe-based pnictides versus formal oxidation state are reversed: the Fe arsenides in the ucT structure exhibit long-range magnetic ordering but not in the cT structure, whereas the reverse is true for the Co pnictides.\cite{Anand2012a}

Comprehensive studies of the physical properties of ${\rm SrCo_2As_2}$ crystals with the ucT ${\rm ThCr_2Si_2}$-type structure were recently reported.\cite{Pandey2013, Jayasekara2013}  Remarkably, from inelastic neutron scattering measurements ${\rm SrCo_2As_2}$ was found to exhibit dynamic AFM spin fluctuations/correlations\cite{Jayasekara2013} with the same in-plane stripe wave vector [$(\pi,0)$ and $(0,\pi)$] as occurs in the ${\rm (Ca,Sr,Ba,Eu)Fe_2As_2}$ parent compounds,\cite{Johnston2010} and the $c$~axis lattice parameter exhibits {\it negative} thermal expansion from~7 to 300~K.\cite{Pandey2013}  Thus  ${\rm SrCo_2As_2}$ does not exhibit long-range magnetic ordering, in agreement with the above phenomenology of magnetic ordering in 122-type Co pnictides.  In view of the similarities of the strength and wavevector of the AFM fluctuations to those in ${\rm BaFe_2As_2}$ and the absence of long-range AFM order in metallic ${\rm SrCo_2As_2}$, the lack of high-$T_{\rm c}$ superconductivity in ${\rm SrCo_2As_2}$ is a conundrum at present.

The compound \cacoastwo\ that we study in this paper is reported to have a cT ${\rm ThCr_2Si_2}$-type structure with a small $c/a = 2.59$.\cite{Pfisterer1980,Pfisterer1983}   Three groups have reported the physical properties of single crystals of this metallic compound. \cite{Cheng2012, Ying2012, Quirinale2013} Cheng et al.\ reported AFM ordering in ${\rm CaCo_2As_2}$ crystals grown from CoAs self-flux at $T_{\rm N} = 76$~K,\cite{Cheng2012}  whereas Ying et al.\ observed AFM ordering at $T_{\rm N} = 70$~K in similarly grown crystals.\cite{Ying2012}  Both groups suggested on the basis of their respective magnetization data that the AFM structure is collinear A-type.  In this AFM structure, the Co ordered magnetic moments within an $ab$~plane layer are ferromagnetically (FM) aligned along the $c$~axis whereas the moments in adjacent layers are aligned antiferromagnetically with respect to each other.

\begin{figure}
\centering\includegraphics[width=2in]{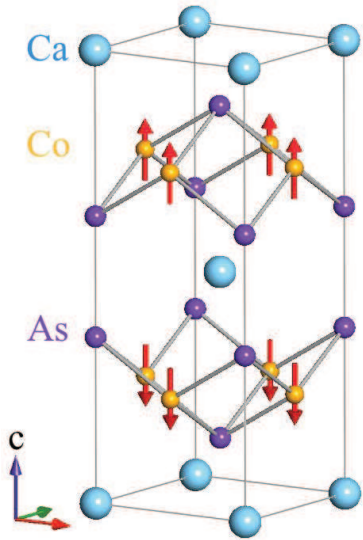}
\caption{(Color online) The body-centered tetragonal magnetic and chemical unit cell of \cacoas\ (space group $I4/mmm$).\cite{Quirinale2013}  The red arrows denote the ordered moment directions along the $c$~axis in the A-type AFM structure determined from neutron diffraction measurements.\cite{Quirinale2013}}
\label{Fig:structure}
\end{figure}

A combined neutron diffraction and synchrotron-based x-ray diffraction (XRD) study of polycrystalline samples of crushed crystals and a laboratory-based x-ray study of a single crystal of \cacoastwo\ were reported by Quiranale et al.,\cite{Quirinale2013} where the crystals were grown in Sn flux instead of CoAs flux.  The results confirmed the ${\rm ThCr_2Si_2}$-type crystal structure as shown in Fig.~\ref{Fig:structure}.  This work further demonstrated that the crystals contain a 7(1)\% vacancy concentration on the Co sites, yielding an actual composition ${\rm CaCo_{1.86(2)}As_2}$, and also contain $\approx 0.5$~mol\% of Sn within the crystals from wavelength-dispersive x-ray analysis.  A similar crystallographic analysis was not done in the two studies of CoAs-grown crystals,\cite{Cheng2012, Ying2012} so it is unknown if those crystals also contain significant concentrations of Co vacancies. 

The work by Quiranale et al.\ also showed that the N\'eel temperature of the \cacoas\ crystals is $T_{\rm N} = 52(1)$~K,\cite{Quirinale2013} significantly lower than the above values of 70--76~K obtained\cite{Cheng2012, Ying2012} for CoAs-grown crystals.  The magnetic structure was found to be A-type as previously inferred\cite{Cheng2012, Ying2012} as shown in Fig.~\ref{Fig:structure}.\cite{Quirinale2013}  The ordered Co moment at 10~K was too small to measure accurately from the neutron diffraction measurements, but an upper limit of $\sim 0.6\,\mu_{\rm B}$/Co was estimated, where $\mu_{\rm B}$ is the Bohr magneton, consistent with the value of $0.3$--$0.4\,\mu_{\rm B}$/Co obtained from magnetization~$M$ versus applied magnetic field~$H$ isotherm measurements at 10~K.\cite{Cheng2012}  The low ordered moment suggests either itinerant AFM or suppression of the $\sim1~\mu_{\rm B}$/Co ordered moments of local Co spins-1/2 due to quantum fluctuations associated with a low effective dimensionality of the Co spin lattice.  From the studies in the present paper, we infer that the former explanation is the correct one.

An investigation of the magnetic phase diagram of CoAs-flux grown crystals of Ca$_{1-x}$Sr$_x{\rm Co_2As_2}$ for \mbox{$0\leq x\leq1$} was carried out by Ying et al.\cite{Ying2013}  From their XRD and $M(H,T)$  measurements, they deduced the occurrence of A-type AFM in the cT phase with the Co ordered moments aligned along the $c$~axis for $0 \leq x \lesssim 0.15$, ferromagnetism in the cT phase with the ordered moments aligned along the $c$~axis for $0.15 \lesssim x \lesssim 0.3$, A-type AFM in the cT phase with moments aligned in the $ab$~plane for $0.3 \lesssim x \lesssim 0.45$, and no long-range magnetic order in the ucT phase for $0.45 \lesssim x \leq 1$.  Thus the above-noted correlation that magnetic ordering occurs in the cT phase but not in the ucT phase of Co pnictide-based ${\rm ThCr_2Si_2}$-type compounds is further confirmed.

Herein we extend the crystallographic and magnetic diffraction studies of Quiranale et al.\cite{Quirinale2013} on our Sn-flux grown \cacoas\ crystals to investigations at room temperature using laboratory-based powder XRD, and at variable temperature~$T$ using magnetic susceptibility $\chi(T)$, isothermal $M(H)$, heat capacity at constant pressure $C_{\rm p}(T)$, electrical resistivity $\rho(T)$ and angle-resolved photoemission spectroscopy (ARPES) measurements.  We also report LDA electronic structure calculations for \cacoastwo\ and for a supercell of the composition ${\rm Ca_8Co_{15}As_{16}}$ (${\rm CaCo_{1.88}As_2}$) that is close to the value for our crystals.

We found from Rietveld refinements of our powder XRD data for crushed ``\cacoastwo'' crystals a Co vacancy concentration of 6.0(5)\%, consistent with the value of 7(1)\% obtained in Ref.~\onlinecite{Quirinale2013} corresponding to an actual composition \cacoas.  Our magnetic susceptibility results are consistent with collinear A-type antiferromagnetic ordering with the ordered moments aligned along the $c$~axis as suggested by Cheng et al.\cite{Cheng2012} and Ying et al.\cite{Ying2012} and confirmed using neutron diffraction by Quirinale et al.\cite{Quirinale2013} (Fig.~\ref{Fig:structure}).  The $T_{\rm N} =52(1)$~K that we and Quirinale et al.\ consistently observe is about 20~K lower than reported in the two recent reports,\cite{Cheng2012, Ying2012} for unknown reasons that may be associated with the different fluxes and/or heating/cooling profiles used for the crystal growths.

The $\rho(T)$, $C_{\rm p}(T)$ and ARPES measurements and the electronic structure calculations consistently indicate that \cacoas\ is metallic.  Our electronic structure calculations and the large value of the Sommerfield coefficient obtained from our $C_{\rm p}(T)$ data both reveal a rather large density of states at the Fermi energy.  Within an itinerant model, we find that \cacoas\ is strongly exchange-enhanced and on the verge of ferromagnetism, consistent with the FM Weiss temperature in the Curie-Weiss law and the FM in-plane alignment of the Co moments in the A-type AFM structure below $T_{\rm N}$.  The band calculations also showed that the FM and A-type AFM structures are nearly degenerate in energy in \cacoastwo.  We modeled our $M(H)$ and $\chi(T)$ data for \cacoas\ using molecular field theory assuming a local moment Heisenberg model and obtained values for the exchange and anisotropy parameters.  However, the low saturation moment of $\approx 0.3~\mu_{\rm B}$/Co obtained from our from $M(H)$ data at low~$T$ and the lack of self-consistent analyses of the entirety of our $M(H)$ and $\chi(T)$ data in terms of the local moment model indicate that \cacoas\ is an itinerant antiferromagnet.  

In Sec.~\ref{ExpDetails} we give details of the experimental and theoretical work.  The crystallography is presented in Sec.~\ref{Crystallography}.  The $M(H)$ and $\chi(T)$ data are presented in Sec.~\ref{Sec:CaCo2As2ChiMH} together with a detailed analysis of these data in terms of both an itinerant picture and molecular field theory for Heisenberg interactions between local magnetic moments.  The $C_{\rm p}(T)$ data are presented in Sec.~\ref{Sec:CaCo2As2_HC}, where the only indication of magnetic ordering at $\sim50$~K is a possible small change in slope, and the $\rho(T)$ data in Sec.~\ref{Sec:CaCo2As2Rho} also show little indication of the AFM ordering.  The ARPES measurements in Sec.~\ref{Sec:CaCo2As2ARPES} indicate little change in the Fermi surface topology at $T_{\rm N}$, although there is some indication that the bands are somewhat temperature dependent overall.  The electronic structure calculations presented in Sec.~\ref{Sec:CaCo2As2DOS} show a sharp and high peak in the density of states (DOS) at or near the Fermi energy $E_{\rm F}$  that arises from the Co $3d$ bands, consistent with the large density of states inferred from the $C_{\rm p}(T)$ data.  This section also contains spin-polarized calculations of the band structure and total energy of the FM and several AFM structures.  A summary of the paper is given in Sec.~\ref{Conclusions}.

\section{\label{ExpDetails} Experimental and Theoretical Details}

Single crystals of \cacoas\ were grown using Sn flux and their composition analyzed as described in detail in Ref.~\onlinecite{Quirinale2013}. Powder XRD measurements using Cu~K$_\alpha$ radiation were carried out with a Rigaku Geigerflex x-ray diffractometer. A Quantum Design, Inc., superconducting quantum interference device magnetic properties measurement system (MPMS) was used for the $M$ and $\chi$ measurements, where the contribution from the empty sample holder was corrected for except for measurements in low (0.01~T) fields where the contribution is negligible.  A Quantum Design, Inc.\ physical properties measurement system (PPMS) was used for $C_{\rm p}$ and $\rho$ measurements. The $C_{\rm p}$ was measured by the relaxation method and the $\rho$ was measured using the standard four-probe ac technique. High magnetic field isothermal $M(H)$ data were collected using the vibrating sample magnetometer (VSM) option of the PPMS, and the contribution of the empty sample holder was subtracted from the measured data.

ARPES experiments were performed using a laboratory-based system consisting of a Scienta SES2002 electron analyzer, GammaData UV lamp and custom-designed refocusing optics at the Ames Laboratory. The samples were cooled using a closed-cycle refrigerator and cleaved in-situ yielding flat, shiny mirror-like surfaces. All data were acquired using the He~I line with a photon energy of 21.2~eV and in ultrahigh vacuum below $5 \times 10^{-11}$~torr. The total experimental energy and angle resolutions were set at $\leq20$~meV and $\leq0.3^\circ$, respectively. The Fermi energy ($E_{\rm F}$) of the sample was referenced to that of a Au sample deposited {\it in situ} on the sample holder.

The full-potential linearized augmented plane-wave (FPLAPW) method \cite{Blaha2001} with the local density approximation (LDA) \cite{Perdew1992} was used for electronic structure calculations. To obtain a self-consistent charge density, we employed 2.1~a.u.\ muffin-tin (MT) radii for all atoms and $R_{\rm MT}k_{\rm max} = 8.0$, where $R_{\rm MT}$ is the smallest muffin-tin radius and $k_{\rm max}$ is the maximum wavevector value in the expansion of plane waves. Calculations were iterated to reach the total energy convergence criterion that was 0.01~mRy/f.u., where f.u.\ means formula unit.  The experimental lattice constants $a=b=3.983,\  c=10.273$~\AA\ from Table~\ref{tab:XRD1} below were used, and the As atom position was relaxed until the force on the As atom was less than 0.01~mRy/a.u. For the nonmagnetic (paramagnetic, PM) calculation, this gave $z_{\rm As} = 0.3622$ as the theoretical As $c$~axis coordinate, which is similar to the experimental value $z_{\rm As} = 0.3672(2)$ in Table~\ref{tab:XRD2} below.

\section{\label {Crystallography} Crystallography}

\begin{figure}
\includegraphics[width=3.3in]{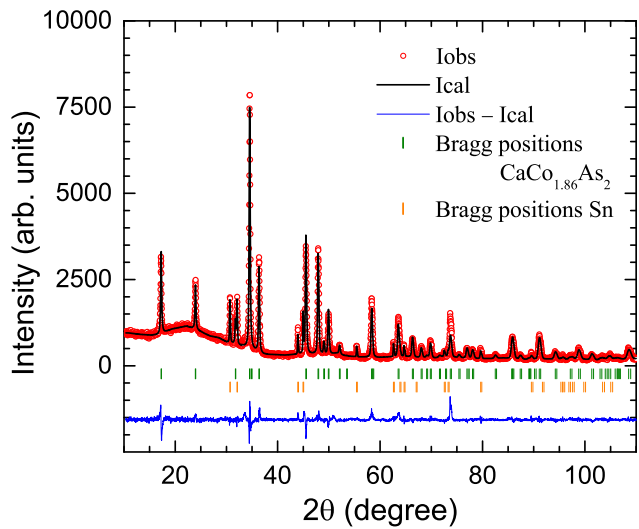}
\caption{ (Color online) Powder x-ray diffraction pattern of \cacoas\ recorded at room temperature. The solid line through the experimental points is the two-phase Rietveld refinement profile calculated for the body-centered tetragonal (bct) ThCr$_2$Si$_2$-type structure (space group $I4/mmm$), including the influence of vacancies on the Co site, and the bct $\beta$-Sn  (space group $I4_1/amd$) structure.  The short vertical bars mark the fitted Bragg peak positions. The lowermost curve represents the difference between the experimental and calculated intensities.}
\label{fig:CaCo2As2_XRD}
\end{figure}

\begin{table}
\caption{\label{tab:XRD1} Crystallographic and Rietveld refinement parameters obtained from room-temperature powder XRD data of crushed \cacoas\ crystals.}
\begin{ruledtabular}
\begin{tabular}{ll}
Structure & ${\rm ThCr_2Si_2}$-type\\
Space group & $I4/mmm$ \\
Formula units/unit cell & $Z = 2$\\
\underline{Lattice parameters}  \\
\hspace{0.8 cm}$a$ (\AA)     &  3.9837(2) \\
\hspace{0.8 cm}$c$ (\AA)     &  10.2733(4) \\
\hspace{0.8 cm}$c/a$         &  2.5788(3) \\
\hspace{0.8 cm}$V_{\rm cell}$ (\AA$^3$) & 163.04(1) \\
\underline{Refinement quality} \\
\hspace{0.8 cm}    $\chi^2$	   & 3.90  \\	
\hspace{0.8 cm}    $R_{\rm p}$ (\%)  & 6.02   \\
\hspace{0.8 cm}    $R_{\rm wp}$ (\%)  & 8.45  \\
\end{tabular}
\end{ruledtabular}
\end{table}

\begin{table}
\caption{\label{tab:XRD2} Atomic coordinates obtained from the Rietveld refinement of room-temperature powder XRD data of crushed \cacoas\ crystals.}
\begin{ruledtabular}
\begin{tabular}{ccccccc}
  Atom & Wyckoff   &	 $x$ 	&	$y$	&	$z$	 & Fractional \\	
   	& position 		& 			&		& 		& occupancy\\
   	& 				& 			&		& 		& (\%)\\
\hline
    Ca & $2a$  	&	 0 	&	0	&   0		& 100  \\
    Co & $4d$	    &	 0 	&	1/2	&	1/4 	& 94.0(5)  \\
    As & $4e$ 	&	 0 	&   0 	&	0.3672(2) & 100 \\
\end{tabular}
\end{ruledtabular}
\end{table}

The room-temperature powder XRD data collected on crushed \cacoas\ single crystals were analyzed by Rietveld refinement using the {\tt FullProf} software. \cite{Rodriguez1993} The XRD pattern and the Rietveld refinement profile are shown in Fig.~\ref{fig:CaCo2As2_XRD}. The crystal data and refinement parameters are listed in Tables~\ref{tab:XRD1} and \ref{tab:XRD2} and the former are in good agreement with previous values.\cite{Pfisterer1980, Pfisterer1983, Quirinale2013}  The refinement revealed the presence of a significant amount of adventitious Sn impurity phase from the growth flux that could not be removed from the crystal surfaces before crushing them and which was accounted for using a two-phase refinement as shown in Fig.~\ref{fig:CaCo2As2_XRD}.

During the refinement the thermal parameters $B$ were kept fixed to $B \equiv 0$ since within the error bars the lattice parameters and the As $c$~axis position parameter $z_{\rm As}$ were insensitive to changes in $B$\@. However, we found that the refinement quality and the calculated line intensities significantly depended on the fractional occupancy of the Co~$4d$ sites. Therefore we refined the occupancy of Co keeping the Ca and As occupancies fixed to the stoichiometric values of unity, and thus obtained an occupancy of 94.0(5)\% for Co. This indicates a vacancy concentration on the Co sites of $6.0(5)$\% corresponding to the composition ${\rm CaCo_{1.88}As_2}$, in agreement with the Co vacancy concentration of 7(1)\% found in Ref.~\onlinecite{Quirinale2013}.  In this paper we cite and use the composition ${\rm CaCo_{1.86}As_2}$ determined for our crystals in Ref.~\onlinecite{Quirinale2013}. 

We obtained the $c/a$ ratio and the interlayer As--As distance $d_{\rm As-As} = (1-2z_{\rm As})c $ using the crystallographic data in Tables~\ref{tab:XRD1} and \ref{tab:XRD2}, yielding $c/a=2.5788(3)$ and $d_{\rm As-As} = 2.729(5)$~\AA\@. These values indicate that \cacoas\ is in the cT phase\cite{Anand2012a} in agreement with previous reports.\cite{Pfisterer1980, Pfisterer1983, Quirinale2013}

\section{\label{Sec:CaCo2As2ChiMH} Magnetization and Magnetic Susceptibility}

\begin{figure}
\includegraphics[width=3.3in]{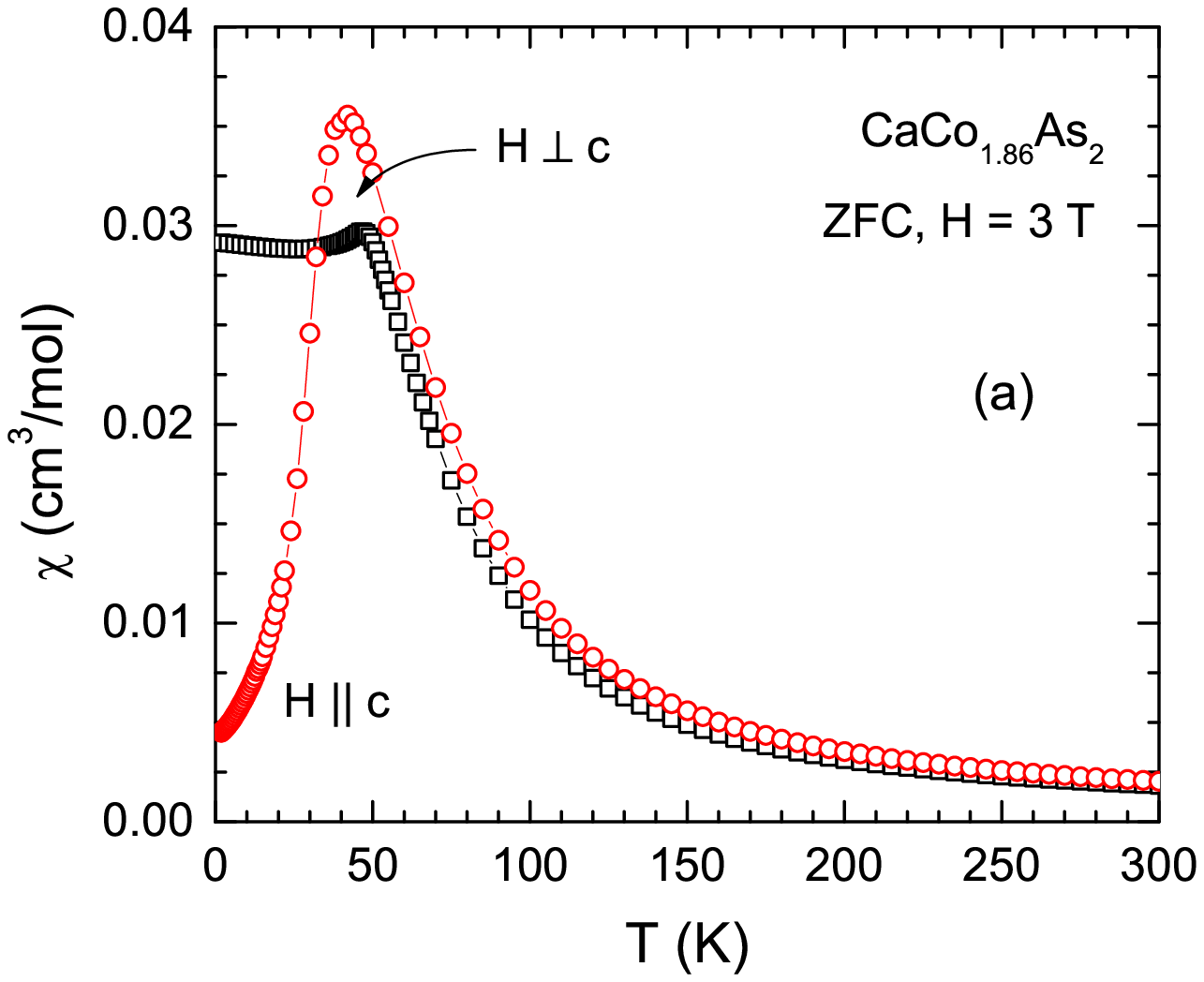}\vspace{0.1in}
\includegraphics[width=3.3in]{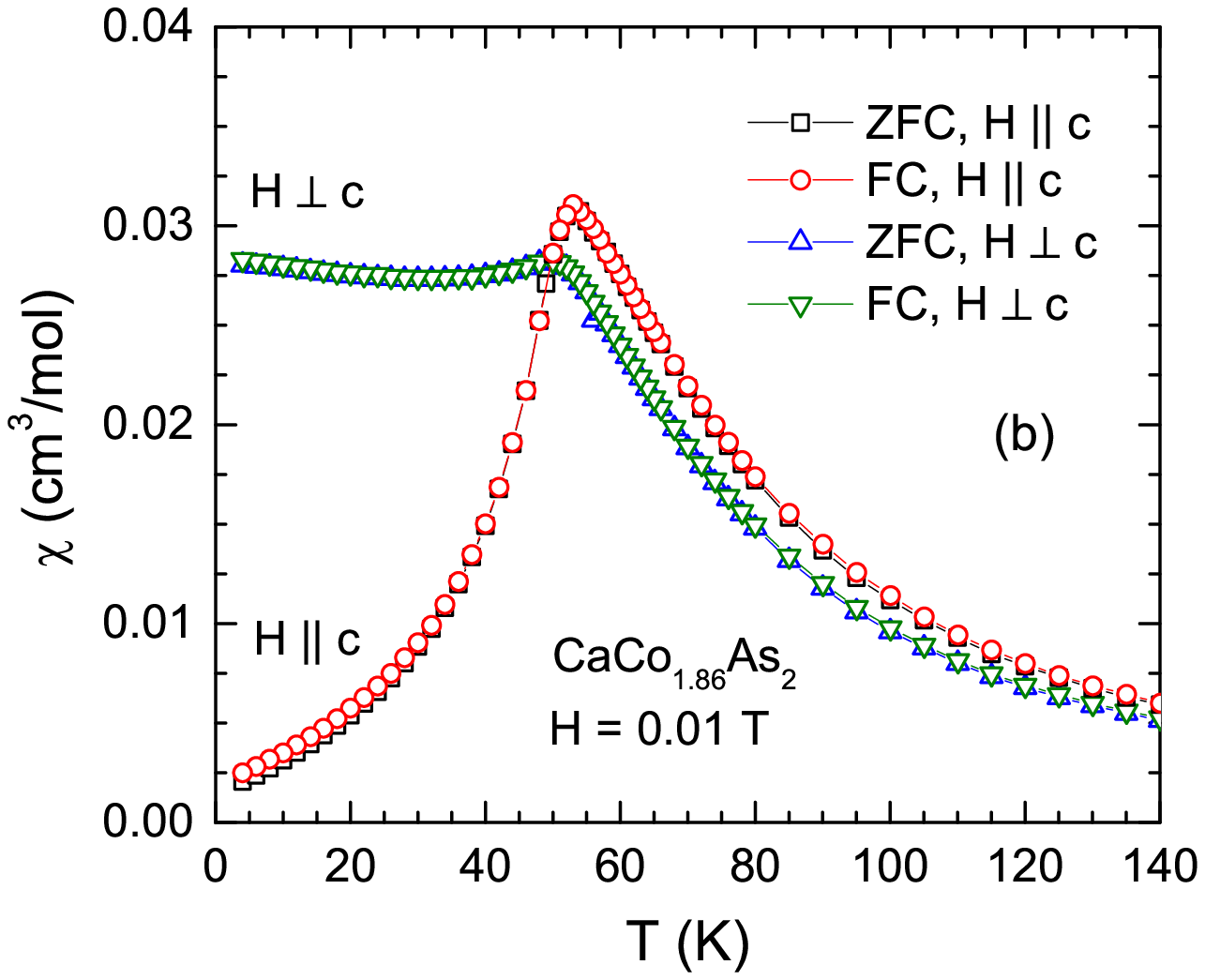}
\caption{(Color online) Zero-field-cooled (ZFC) and field-cooled (FC) magnetic susceptibility $\chi$ of a \cacoas\ single crystal as a function of temperature $T$ in the temperature range 1.8--300~K measured in a magnetic field (a) $H = 3.0$~T and (b) $H = 0.01$~T applied along the $c$~axis ($\chi_c, H \parallel c$) and in the $ab$~plane ($\chi_{ab}, H \perp  c$).}
\label{fig:MT_CaCo2As2_low-H}
\end{figure}

\begin{figure}
\includegraphics[width=3.3in]{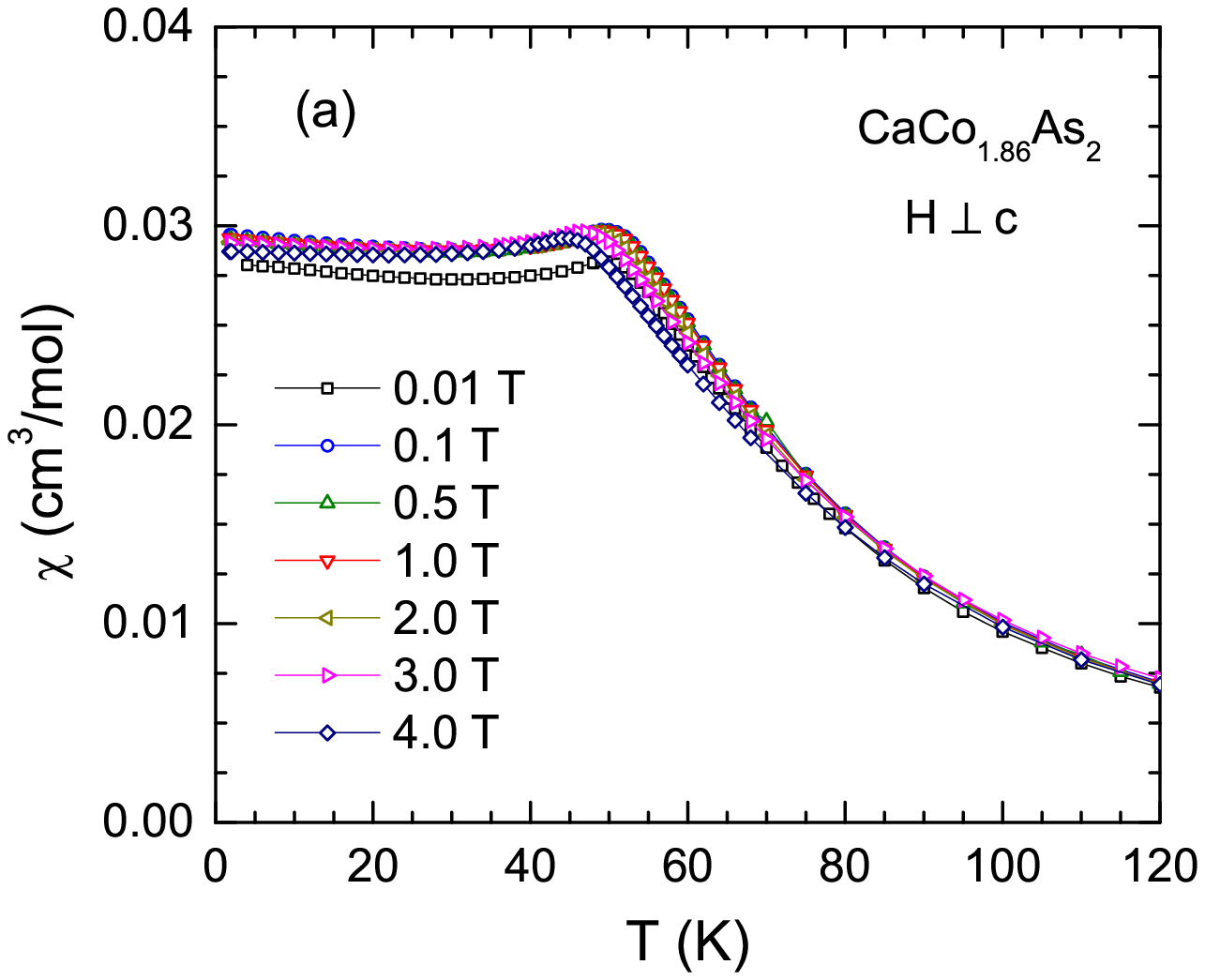}\vspace{0.1in}
\includegraphics[width=3.3in]{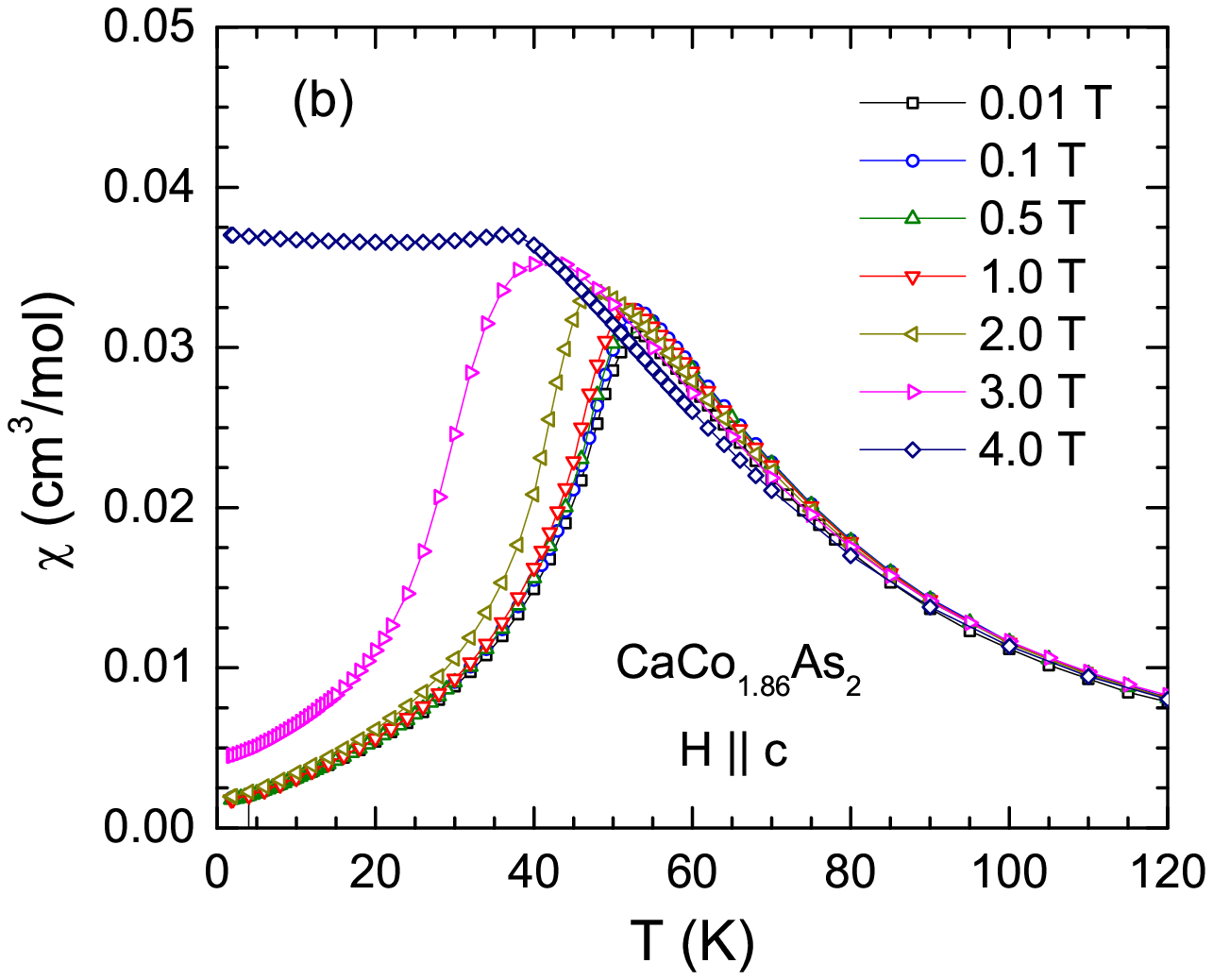}
\caption{(Color online) Zero-field-cooled magnetic susceptibility $\chi$ of a \cacoas\ single crystal as a function of temperature $T$ in the temperature range 1.8--120~K measured in different magnetic fields $H$ applied (a) in the $ab$~plane ($\chi_{ab}, H \perp  c$) and (b) along the $c$~axis ($\chi_c, H \parallel c$).}
\label{fig:MT_CaCo2As2}
\end{figure}

The $\chi \equiv M/H$ data of a \cacoas\ single crystal as a function of $T$ measured at different $H$ applied along the $c$~axis ($\chi_c,\ H \parallel c$) and in the $ab$~plane ($\chi_{ab},\  H \perp c$) are shown in Figs.~\ref{fig:MT_CaCo2As2_low-H} and \ref{fig:MT_CaCo2As2}. At low fields, the $T$ dependence of $\chi$ for $H\parallel c$ in Fig.~\ref{fig:MT_CaCo2As2_low-H}(b) exhibits a well-defined cusp at $T_{\rm N} = 52(1)$~K typical of an AFM phase transition.  From Figs.~\ref{fig:MT_CaCo2As2_low-H}(b) and \ref{fig:MT_CaCo2As2}(b), when $H$ is applied along the $c$~axis the $T_{\rm N}$ decreases from 52 to 42~K as the field increases to $H = 3.0$~T; such a decrease is expected for an AFM transition.  On the other hand, for $H$ applied in the $ab$~plane the transition in $\chi(T)$ is less well-defined.

In order to obtain more insight about the nature of the transition we also collected zero-field-cooled (ZFC) and field-cooled (FC) $\chi(T)$ data at $H = 0.01$~T\@ which are shown in Fig.~\ref{fig:MT_CaCo2As2_low-H}(b). No hysteresis is observed between the ZFC and FC data, which is consistent with long-range AFM ordering of \cacoas\ below $T_{\rm N}$.

From Fig.~\ref{fig:MT_CaCo2As2_low-H}, the anisotropy in $\chi$ for $T < T_{\rm N}$ indicates that  the AFM structure is collinear with the easy axis along the $c$~axis. This observation of ordering along the $c$~axis contrasts with the AFM stripe ordering with the ordered moment oriented in the $ab$~plane in the iron arsenide family, e.g., in $A{\rm Fe_2As_2}$ and ${\rm EuFe_2As_2}$,\cite{Johnston2010} but is the same as the easy axis for G-type AFM ordering of the Mn spins~$S=5/2$ in ${\rm BaMn_2As_2}$.\cite{SinghY2009}

\begin{figure}
\includegraphics[width=3.3in]{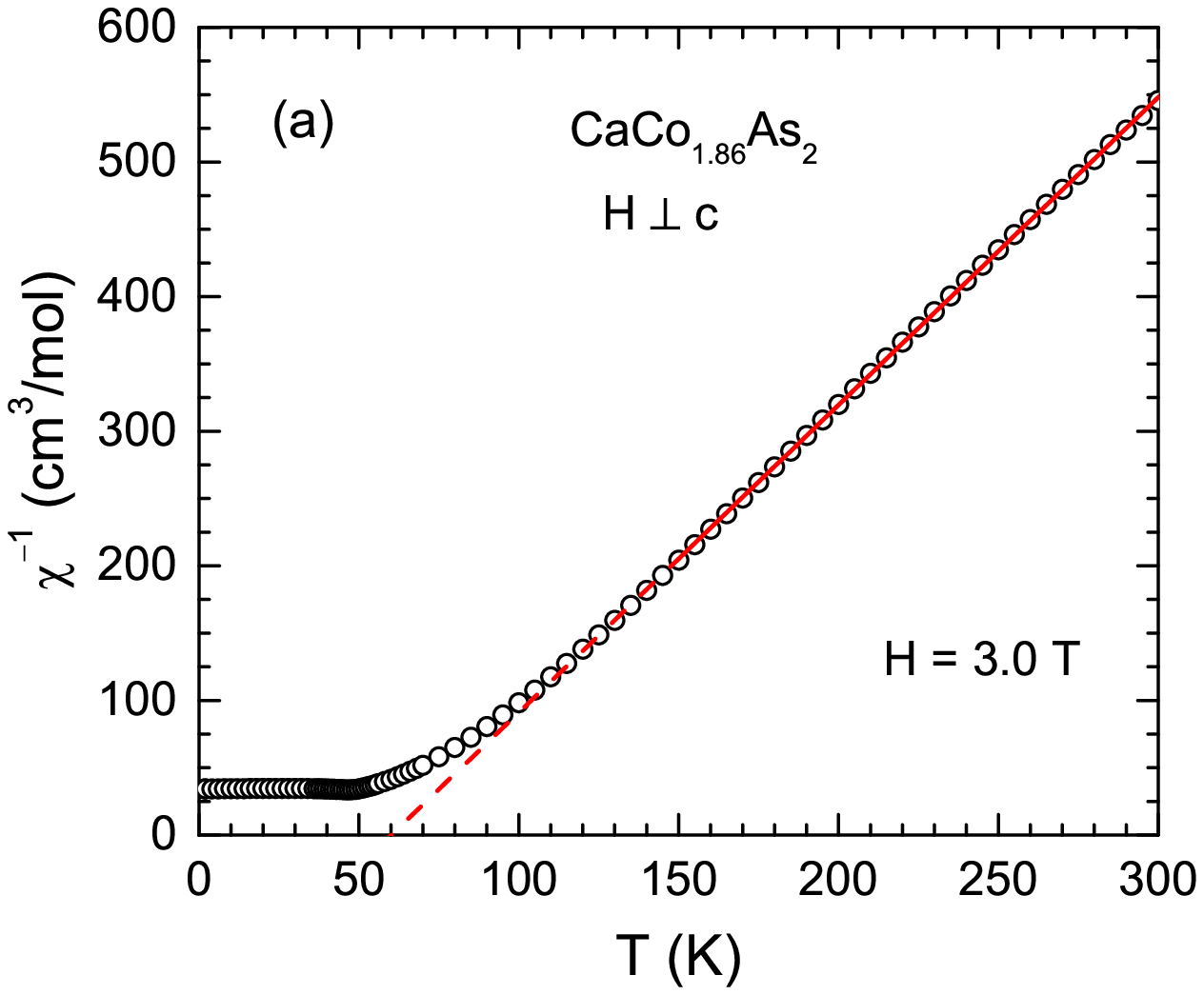}\vspace{0.1in}
\includegraphics[width=3.3in]{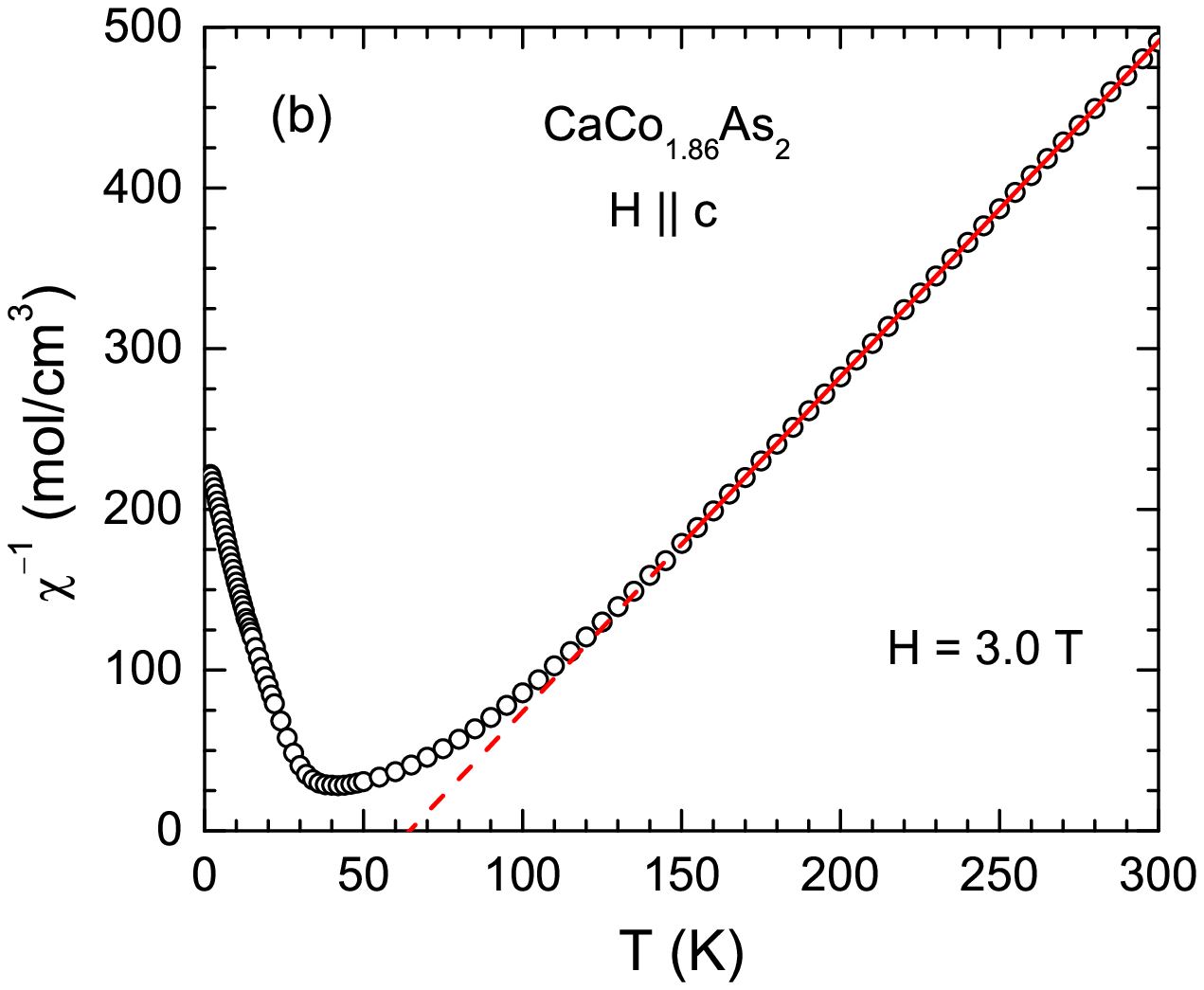}
\caption{(Color online) Zero-field-cooled inverse magnetic susceptibility $\chi^{-1}$ of a \cacoas\ single crystal as a function of temperature $T$ in the temperature range 1.8--300~K measured in a magnetic field $H=3.0$~T applied (a) in the $ab$~plane ($\chi_{ab}, H \perp  c$) and (b) along the $c$~axis ($\chi_c, H \parallel c$). The solid straight red lines are the respective fits of $\chi^{-1}(T)$ data to the Curie-Weiss (CW) law in Eq.~(\ref{eq:C-W}) in the $T$~range 150~K~$\leq T \leq$~300~K\@. The dashed straight red lines are the extrapolations of the respective CW fits to lower temperatures.}
\label{fig:MT_CaCo2As2_inv}
\end{figure}

In the paramagnetic state at sufficiently high temperatures above $T_{\rm N}$, $\chi(T)$ follows Curie-Weiss behavior
\begin{equation}
\chi(T) = \frac{C}{T-\theta_{\rm p}}
\label{eq:C-W}
\end{equation}
where $C$ is the Curie constant and $\theta_{\rm p}$ is the Weiss temperature. The anisotropic $\chi^{-1}(T)$ data measured at $H=3$~T are shown in Fig.~\ref{fig:MT_CaCo2As2_inv} together with the respective fits by Eq.~(\ref{eq:C-W}) for 150~K~$\leq T \leq$~300~K\@. The fitting parameters are $C = 0.480(1)$~cm$^3$\,K/mol and $\theta_{\rm p}^c= +64.5(3)$~K for $\chi_c$ and $C = 0.443(3)$~cm$^3$\,K/mol and $\theta_{\rm p}^{ab} = +59.5(9)$~K for $\chi_{ab}$, which are  also listed in Table~\ref{tab:CW} and compared with literature data.\cite{Cheng2012, Ying2012} A Co atom in the Co$^{+1}$ state with $3d^8$ electronic configuration is expected to have spin $S=1$. For spins with gyromagnetic factor $g$ the Curie constant is $C = [0.5002\,g^2S(S+1)/4]~{\rm cm^3\,K}$ per mole of spins.  Thus for $S=1$ and $g = 2$ one expects $C = 1.86~{\rm cm^3\,K/mol}$ for \cacoas\ which is about a factor of four larger than the experimental values.  For $S = 1/2$, one calculates $C = 0.70~{\rm cm^3\,K/mol}$ for \cacoas\ with $g=2$, which is closer to the observed values but still about 50\% larger. One can obtain the observed $C$ values if 
\be
S = 1/2\quad {\rm and}\quad g \approx 1.6.
\label{Eq:S12g}
\ee
A low-spin $S=1/2$ state for Co can be obtained if the formal oxidation state of the Co is Co$^{+2}$ with a $3d^7$ electronic configuration.  However the $g$ value is more difficult to explain in a local-moment picture.  \cacoas\ is not an ionic insulator and we conclude below that the magnetism is more likely itinerant rather than arising from local moments.

\begin{table}
\caption{\label{tab:CW} AFM ordering temperature $T_{\rm N}$ and the parameters obtained from Curie-Weiss fits of the magnetic susceptibility of CaCo${_{2-x}\rm As_2}$ single crystals, including data from the literature.  ``PW'' means present work.  $f$ is defined as the ratio $f \equiv \theta_{\rm p}/T_{\rm N}$.  The relationship between the effective moment $\mu_{\rm eff}$ per Co assuming $g=2$ and the Curie constant $C$ is $\mu_{\rm eff} = \sqrt{8C/(2-x)}$ for $C$ and $\mu_{\rm eff}$ expressed in the units given, where $x=0.14$ for our crystals and we set $x=0$ for the literature values.}
\begin{ruledtabular}
\begin{tabular}{ccccccc}
Field   & $T_{\rm N}$  & $C$  &  $\theta_{\rm p}$ & $f$  &	$\mu_{\rm eff}$ & Ref.\\
direction & (K) & (${\rm \frac{cm^3\,K}{mol}}$) & (K) & & ($\mu_{\rm B}$/Co) \\
\hline
$H \parallel c$  & 53 &  0.480(1)  & +64.5(3) &  1.22  &1.41(1) & PW  \\
 & 76  &  & 65 & &1.4 & [\onlinecite{Cheng2012}]\\	
 &   &  & 63 & &1.15 & [\onlinecite{Ying2012}]\\
$H \perp c$      & 50 &  0.443(3)  & +59.5(9) & 1.19 &1.35(1) & PW  \\	
 & 76 & & 98 & &1.0 & [\onlinecite{Cheng2012}]\\	
 & 70  &  & 91 & & 0.9 & [\onlinecite{Ying2012}]\\		
\end{tabular}
\end{ruledtabular}
\end{table}

\begin{figure}
\includegraphics[width=3.3in]{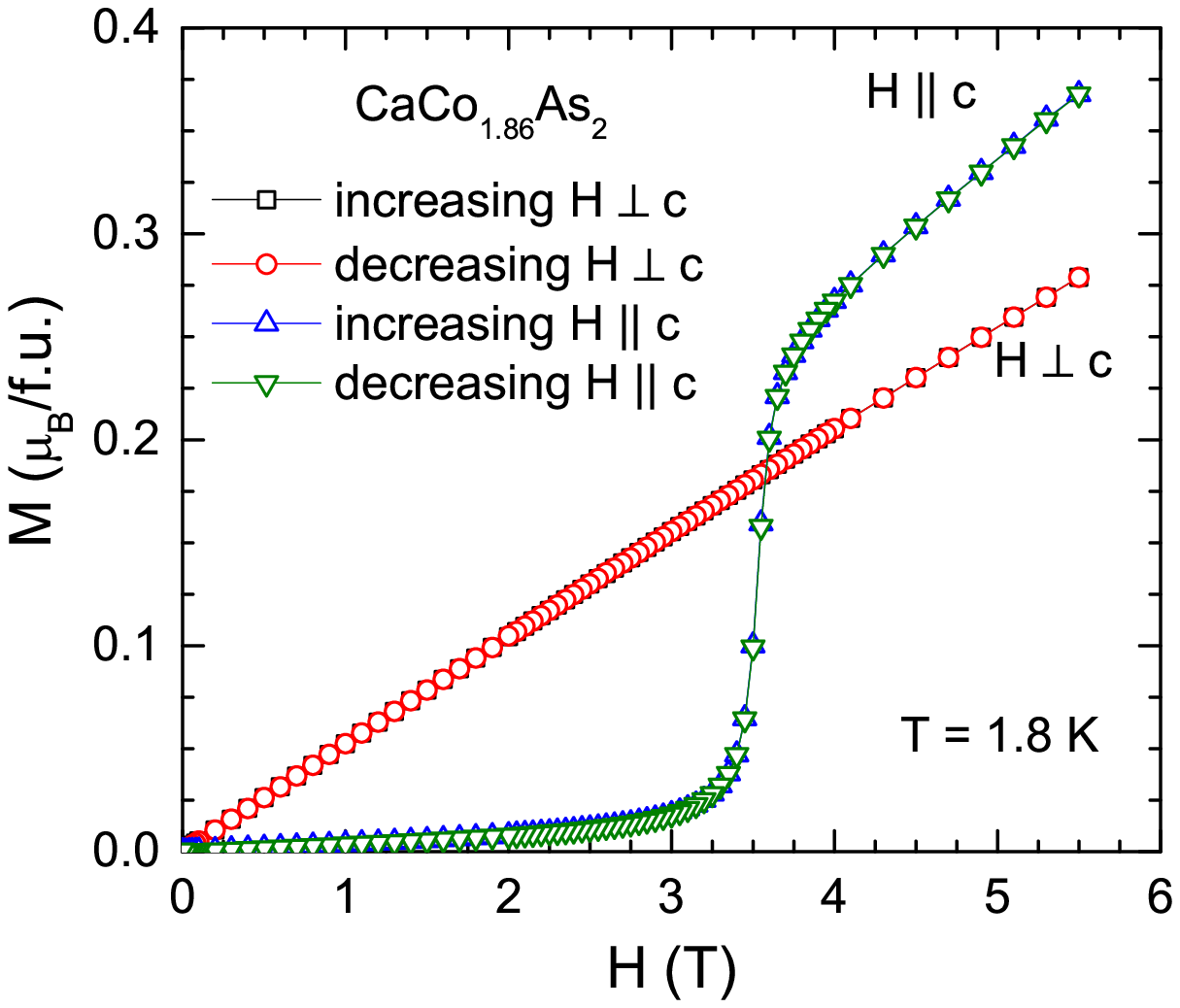}
\caption{(Color online) Isothermal magnetization $M$ of a \cacoas\ single crystal as a function of applied magnetic field $H$ measured at 1.8~K for $H$ applied in the $ab$~plane ($M_{ab}, H \perp  c$) and along the $c$~axis ($M_c, H \parallel c$).}
\label{fig:MH_CaCo2As2_2K}
\end{figure}

\begin{figure}
\includegraphics[width=3.3in]{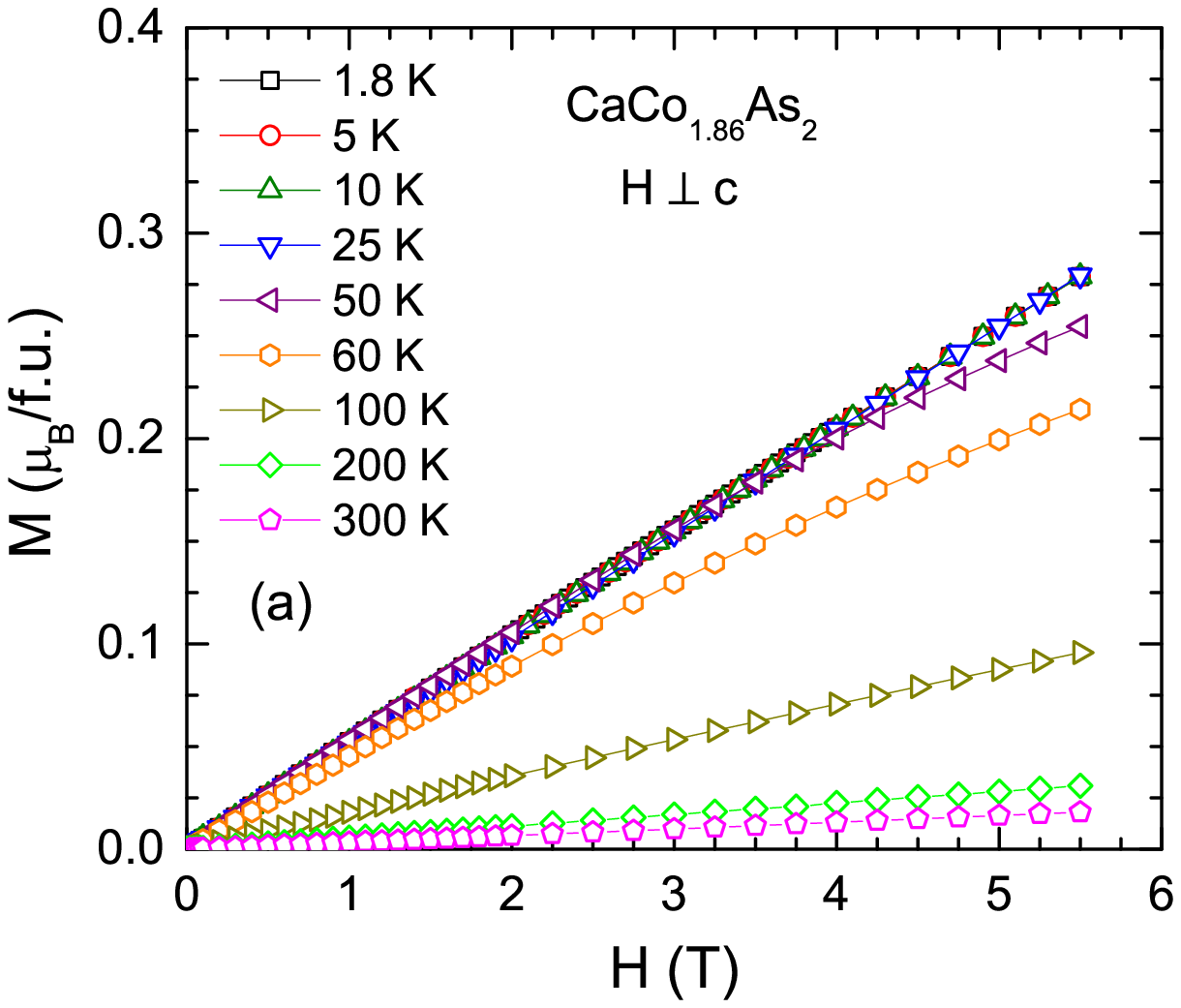}\vspace{0.1in}
\includegraphics[width=3.3in]{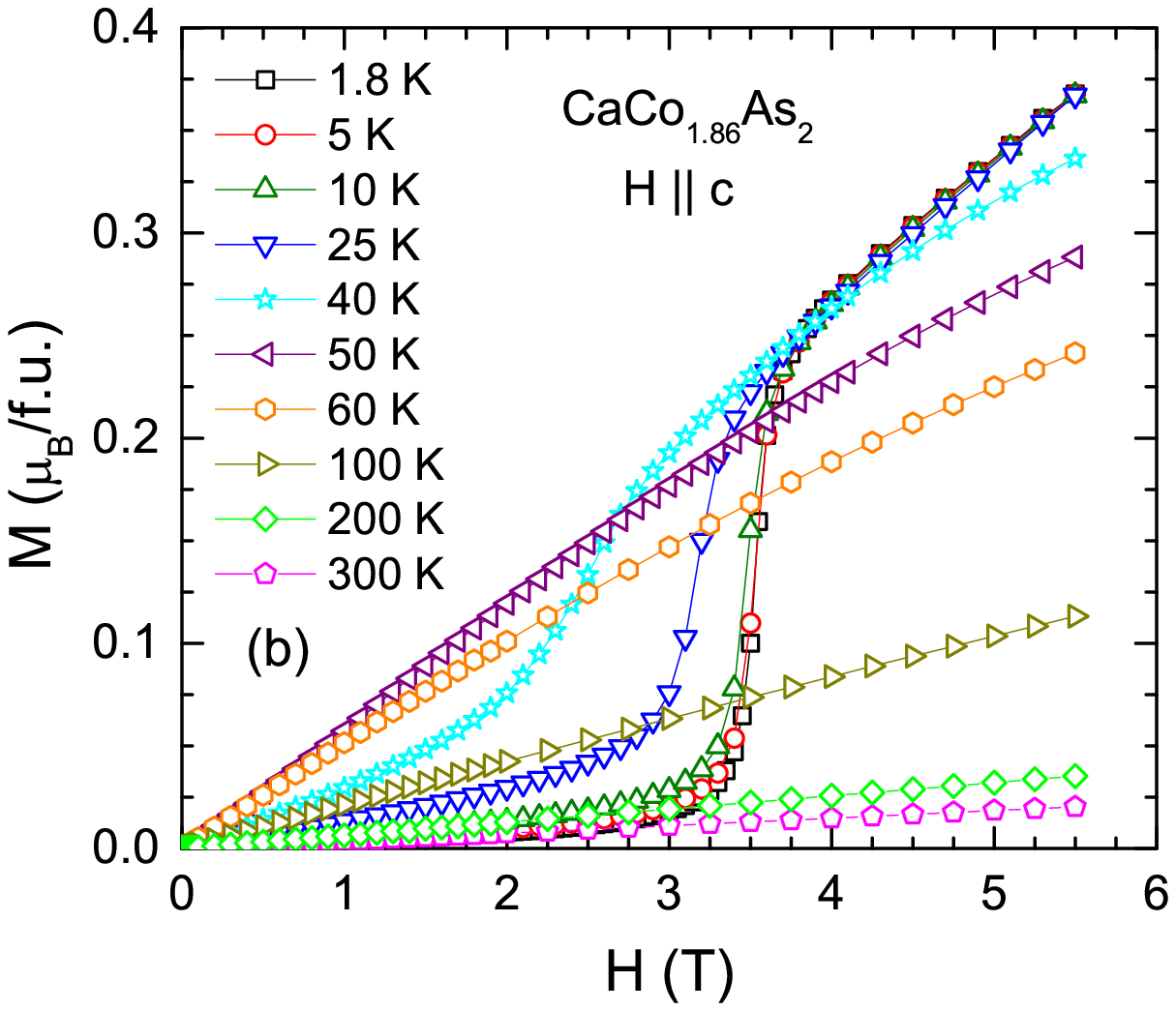}
\caption{(Color online) Isothermal magnetization $M$ of a \cacoas\ single crystal as a function of applied magnetic field $H$ measured at the indicated temperatures for $H$ applied (a) in the $ab$~plane ($M_{ab}, H \perp  c$) and (b) along the $c$~axis ($M_c, H \parallel c$).}
\label{fig:MH_CaCo2As2}
\end{figure}

The isothermal $M(H)$ behaviors of a ${\rm CaCo_{1.86}As_2}$ crystal at different temperatures between 1.8~K and 300~K for $H$ applied along the $c$~axis ($M_c, H \parallel c$) and in the $ab$~plane ($M_{ab}, H \perp  c$) are shown in Figs.~\ref{fig:MH_CaCo2As2_2K} and \ref{fig:MH_CaCo2As2}. At 1.8~K, the $M_{ab}(H)$ data in Fig.~\ref{fig:MH_CaCo2As2_2K} are almost proportional to $H$, whereas the $M_c(H)$ isotherm exhibits a sharp jump due to a field-induced spin-flop transition at $H_{\rm Flop} = 3.5$~T as determined from the field at which $dM_c(H)/dH$ is maximum.  No hysteresis is observed in Fig.~\ref{fig:MH_CaCo2As2_2K} between the $M(H)$ isotherms at 1.8~K upon increasing and decreasing $H$ for either $H \parallel c$ or $H \perp  c$.

The $M$ at 1.8~K does not saturate up to 5.5~T for either $H \parallel c$ or $H \perp  c$ (Fig.~\ref{fig:MH_CaCo2As2_2K}). As $T$ is increased, $M$ versus~$H$ for $H \perp  c$ shows negative curvature  at 60~K and 50~K but is nearly proportional for $T \leq 25$~K and $T \geq 100$~K as shown in Fig.~\ref{fig:MH_CaCo2As2}.  On the other hand the $M(H)$ isotherms measured at different $T$ for $H \parallel c$ exhibit clear spin-flop transitions for $T \leq 40$~K, where $H_{\rm Flop}$ increases with decreasing~$T$\@. 

\begin{figure}
\includegraphics[width=3.3in]{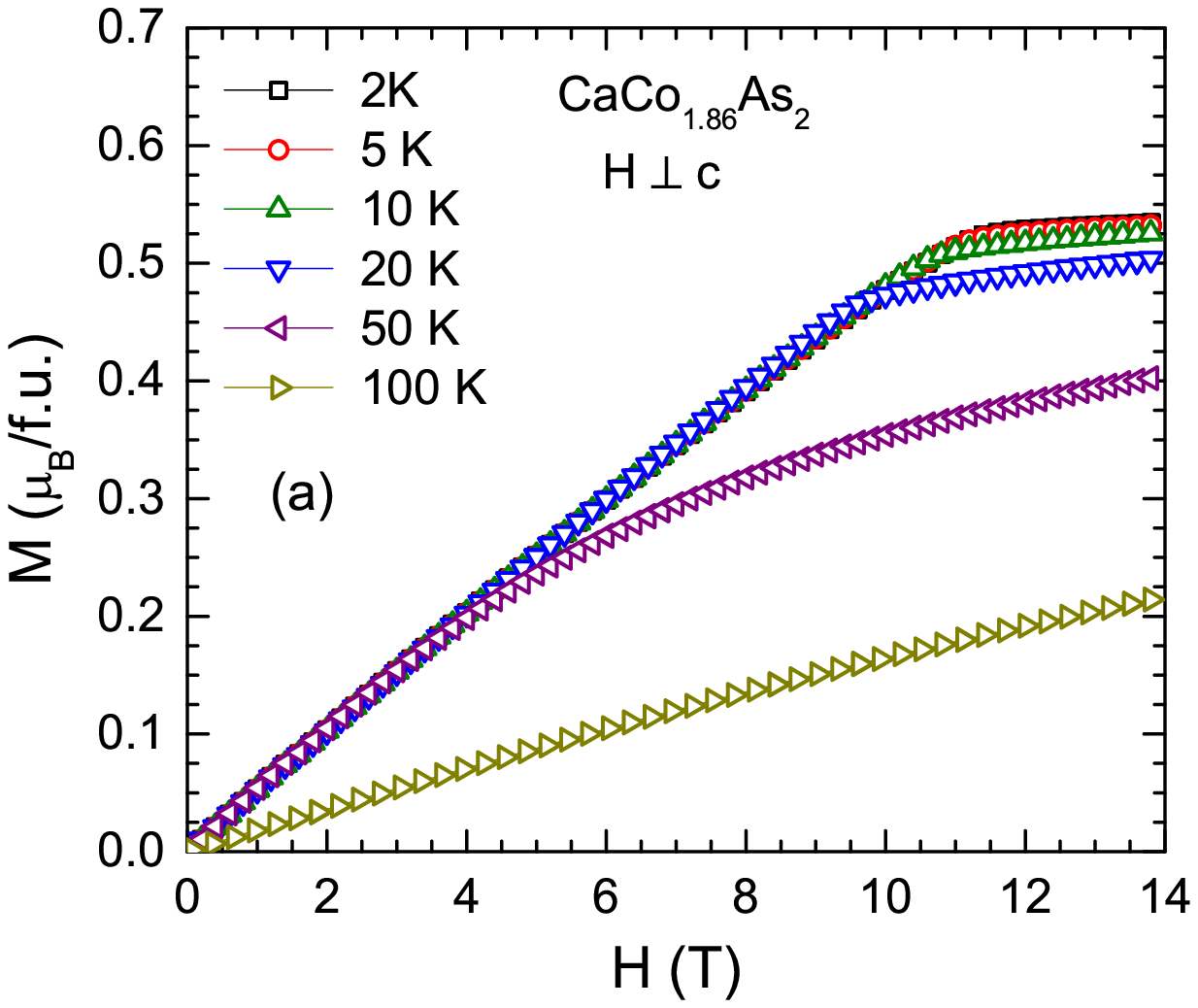}\vspace{0.1in}
\includegraphics[width=3.3in]{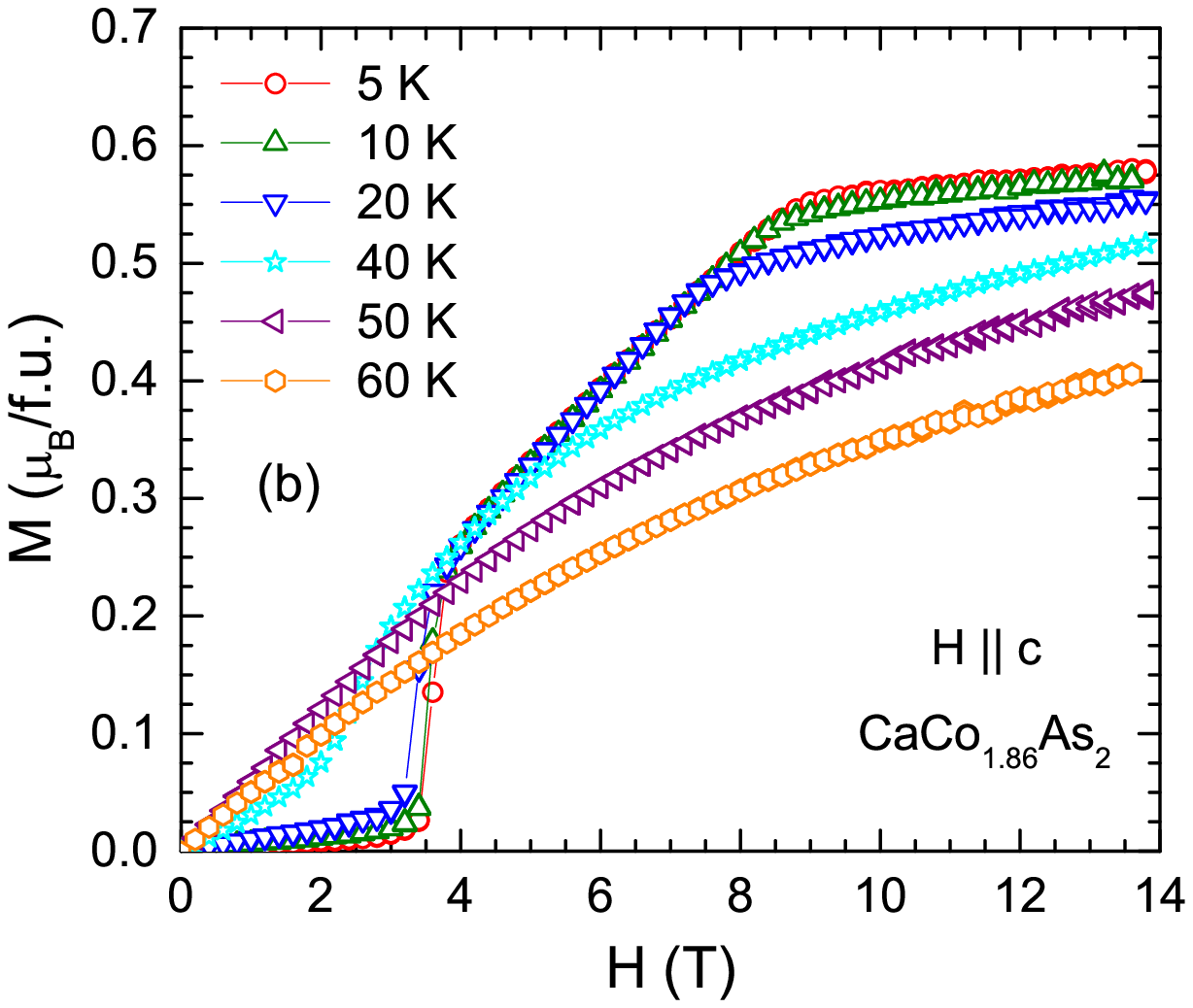}
\caption{(Color online) High-field isothermal magnetization $M$ of a \cacoas\ single crystal as a function of applied magnetic field $H$ at the indicated temperatures for $H$ applied (a) in the $ab$~plane ($M_{ab}, H \perp  c$) and (b) along the $c$~axis ($M_c, H \parallel c$).}
\label{fig:MH_CaCo2As2_VSM}
\end{figure}

\begin{figure}
\includegraphics[width=3.3in]{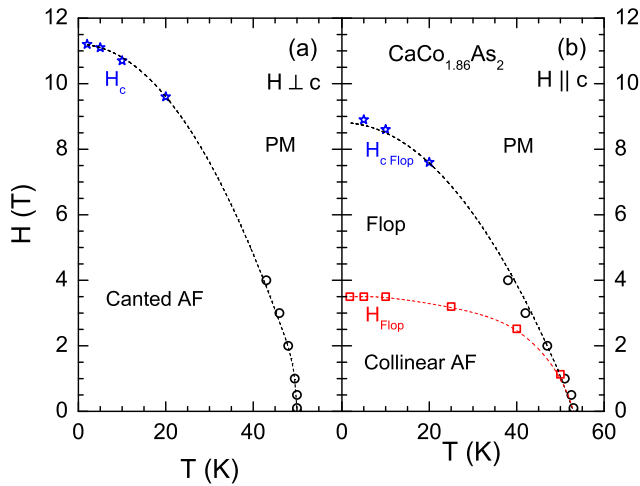}
\caption{(Color online) $H$-$T$ phase diagram for \cacoas\ obtained from magnetic susceptibility $\chi(T)$ and isothermal magnetization  $M(H)$ data for $H$ applied (a) in the $ab$~plane ($H \perp  c$) and, (b) along the $c$~axis ($H \parallel c$). The curves are guides to the eye. }
\label{fig:HT_CaCo2As2}
\end{figure}

$M(H)$ data for the same crystal measured at higher fields up to 14~T are shown in Fig.~\ref{fig:MH_CaCo2As2_VSM}. It is seen from this figure that the saturation moments and the critical fields separating the canted AFM from the paramagnetic (PM) phases are somewhat different for $H \perp  c$ and $H \parallel c$. In particular, for $H \perp  c$ the saturation moment $M_{\rm sat}^{ab} =0.53\,\mu_{\rm B}$/f.u., whereas for $H \parallel c$ one obtains $M_{\rm sat}^{c} =0.58\,\mu_{\rm B}$/f.u. These values correspond to a saturation (ordered) moment $\mu_{\rm sat} = 0.27$ and $0.29~\mu_{\rm B}$/Co, respectively.  Furthermore, the critical field for the transition between the PM and AFM states is found to be $H_{\rm c}Ê= 11.2$~T from the $M(H)$ data in Fig.~\ref{fig:MH_CaCo2As2_VSM}(a) at 2~K and $H_{\rm c\,Flop} = 8.9$~T from the $M(H)$ data in Fig.~\ref{fig:MH_CaCo2As2_VSM}(b) at 5~K, where the critical fields are labeled according to whether the applied field is perpendicular or parallel to the easy $c$~axis, respectively.

The saturation moments are much smaller than the value $\mu_{\rm sat} = g S \mu_{\rm B}/{\rm Co} = 0.8\,\mu_{\rm B}$/Co expected from the parameters $g=1.6$ and $S=1/2$ in Eq.~(\ref{Eq:S12g}) for a Co atom as derived above from the values of the Curie constants. This inconsistency suggests that an itinerant model for the antiferromagnetism is more appropriate for ${\rm CaCo_{1.86}As_2}$ than a local-moment model.

The $H$-$T$ phase diagram determined from the $H$ dependence of $T_{\rm N}$ from $\chi(T)$ measurements in different $H$ and $M(H)$ measurements at different $T$ is shown in Fig.~\ref{fig:HT_CaCo2As2}. For $H \perp  c$, the only  phase boundary that occurs separates the (field-canted) AFM phase from the PM phase at the critical field $H_{\rm c\perp}$.  For $H \parallel c$ there are two phase transition curves, one at $H_{\rm Flop}(T)$ that separates the AFM and spin-flop phases, and the other at the critical field $H_{\rm c\,Flop}(T)$ that separates the spin-flop and PM phases.

The two phase diagrams in Fig.~\ref{fig:HT_CaCo2As2} for the two field directions are consistent with the A-type AFM structure in Fig.~\ref{Fig:structure}.  The fact that $\chi_c$ in Fig.~\ref{fig:MT_CaCo2As2_low-H}(b) drops by 92\% of its value between $T_{\rm N}$ and~1.8~K indicates that the AFM structure is collinear, where the easy axis (the ordered moment axis) is along $c$.  This is consistent with the observation of a spin-flop transition when $H$ is oriented along $c$.  Then considering that the Weiss temperature $\theta_{\rm p}$ in Table~\ref{tab:CW} is strongly positive indicates that the AFM in ${\rm CaCo_2As_2}$ is of A-type (Fig.~\ref{fig:CaCo2As2_XRD}) as inferred in Refs.~\onlinecite{Cheng2012} and~\onlinecite{Ying2012} and confirmed in Ref.~\onlinecite{Quirinale2013}.  A-type AFM has also been found in isostructural ${\rm CaCo_2P_2}$ ($T_{\rm N} = 113$~K) from neutron diffraction measurements\cite{Reehuis1998} with an ordered moment of $0.32(2)\,\mu_{\rm B}$/Co that is similar to our value in \cacoas.  However, the neutron diffraction measurements also revealed that the ordered Co moments in ${\rm CaCo_2P_2}$ are oriented in the $ab$~plane\cite{Reehuis1998} instead of along the $c$~axis as in \cacoas.

\subsection{\label{Sec:ItElThy} Data Analysis Using Itinerant Electron Theory}

The bare Pauli spin susceptibility of degenerate conduction carriers is given by $\chi^{\rm bare} = (g^2/4) \mu_{\rm B}^2 {\cal D}(E_{\rm F})$, where ${\cal D}(E_{\rm F})$ is the bare band structure density of states for both spin directions.  In convenient units one obtains
\be
\chi^{\rm bare}\left[{\rm \frac{cm^3}{mol}}\right] = \frac{g^2}{4}\left(3.233\times10^{-5}\right){\cal D}(E_{\rm F})\left[{\rm \frac{states}{eV\,f.u.}}\right].
\ee
Using $g=2$ and the values of ${\cal D}(E_{\rm F})$ for the PM and A-type AFM states obtained for ${\rm CaCo_{1.88}As_2}$ in Eqs.~(\ref{Eqs:DEFPMAFM}) from our electronic structure calculations below, one obtains
\bse
\bea
\chi^{\rm bare} &=& 2.36 \times 10^{-4}~{\rm \frac{cm^3}{mol}}\quad ({\rm PM~state}),\label{Eq:ChiPauliPM}\\*
\chi^{\rm bare} &=& 0.95 \times 10^{-4}~{\rm \frac{cm^3}{mol}}\quad({\rm A~type~AFM}).\label{Eq:ChiPauliAFM}
\eea
\ese
A comparison of the PM $\chi^{\rm bare}$ value with the experimental $\chi$ values in Figs.~\ref{fig:MT_CaCo2As2_low-H} and~\ref{fig:MT_CaCo2As2_inv} shows that the observed $\chi(T=300\,{\rm K})$ is about 8.6 times larger than the PM value and $\chi(T=T_{\rm N} = 52\,{\rm K})$ is about 120 times larger than the PM value.  Furthermore, in the A-type AFM state, the powder average of the single crystal data in Fig.~\ref{fig:MT_CaCo2As2_low-H}(b) at 1.8~K is $\chi_{\rm ave} \approx 0.019~{\rm cm^3/mol}$, which is about 200 times larger than the AFM $\chi^{\rm bare}$ value in Eq.~(\ref{Eq:ChiPauliAFM}).  These comparisons consistently indicate that within an itinerant magnetism picture, $\chi$ is strongly exchange enhanced in \cacoas.

The exchange-enhanced Stoner susceptibility of an itinerant electron system is given by \cite{Buschow2004}
\begin{equation}
\chi = \frac{\chi^{\rm bare}}{1-I{\cal N}(E_{\rm F})}
\label{eq:Stoner}
\end{equation}
where $\chi^{\rm bare}$ is the bare magnetic spin susceptibility in the absence of Stoner enhancement, $I$ is the on-site Hund coupling energy in eV and ${\cal N}(E_{\rm F})$ is the density of states at the Fermi energy per magnetic atom for one spin direction in states/eV\@. Thus in terms of the density of states ${\cal D}(E_{\rm F})$ per formula unit for both spin directions that is relevant to the heat capacity measurements below, one has ${\cal N}(E_{\rm F})\approx {\cal D}(E_{\rm F})/4$, where the $\approx$ sign is there because there are 1.86 Co atoms per formula unit of \cacoas\ instead of 2.  Taking ${\cal D}(E_{\rm F}) = 7.29$ states/(eV f.u.)\ for both spin directions in the PM state gives ${\cal N}(E_{\rm F})\approx 1.8$~states/Co for one spin direction.  Then using $I \approx 1.0$~eV for Co from Ref.~\onlinecite{Mohn2003} gives $I{\cal N}(E_{\rm F}) \approx 1.8$, which according to Eq.~(\ref{eq:Stoner}) is more than sufficient to drive the system into a FM state.

In order to further compare the exchange enhancements of the measured $\chi$ for different metals it is useful to define the factor $F= I{\cal N}(E_{\rm F})$, so Eq.~(\ref{eq:Stoner}) can be written as
\begin{equation}
\chi = \frac{\chi^{\rm bare}}{1-F}.
\label{eq:Stoner-Z}
\end{equation}
Therefore we obtain
\begin{equation}
F = 1- \frac{\chi^{\rm bare}}{\chi}.
\label{eq:Stoner-Z-Pauli}
\end{equation}
Taking $\chi^{\rm bare}$ to be the calculated bare Pauli susceptibility in the PM state in Eq.~(\ref{Eq:ChiPauliPM}),  substituting this value and the measured value $\chi_c(T=T_{\rm N}) = 3.09 \times 10^{-2}$~cm$^3$/mol from Fig.~\ref{fig:MT_CaCo2As2_low-H}(b) into Eq.~(\ref{eq:Stoner-Z-Pauli}) gives $F = 0.99$, which is larger than the values $F = 0.83$ for elemental Pd and $F = 0.89$ for YFe$_2$Zn$_{20}$.\cite{Jia2007} Using Eq.~(\ref{eq:Stoner-Z-Pauli}) it is not possible to obtain a value $F>1$ since the spin susceptibilities $\chi^{\rm bare}$ and $\chi$ are both positive.

Thus in an itinerant electron description a large Stoner enhancement of $\chi$ is found in \cacoas\ which is sufficient to lead to itinerant electron ferromagnetism, consistent with the positive Weiss temperature in the Curie-Weiss law (Table~\ref{tab:CW}).  \cacoas\ does order ferromagnetically within the $ab$-plane Co layers.  The reason that \cacoas\ does not order ferromagnetically overall is evidently due to a weak AFM coupling between the ferromagnetically-aligned Co layers, leading to the \mbox{A-type} AFM structure in Fig.~\ref{Fig:structure}.

\subsection{\label{Sec:MFA} Data Analysis Using Molecular Field Theory}

\subsubsection{\label{MFTIntro} Introduction}

In this section we use molecular field theory (MFT) for the Heisenberg model with uniaxial anisotropy along the easy $z$-axis and a magnetic field $H$ applied in the $+z$-direction.\cite{Johnston2012b}  All spins are assumed to be identical and crystallographically equivalent.  The average internal energy $E_i$ of an ordered (including magnetic field-induced) magnetic moment $\vec{\mu}_i$ interacting with its neighbors $\vec{\mu}_j$ and with the applied and anisotropy fields, which are both aligned along the $z(c)$ axis, is
\be
E_i = \frac{1}{2g^2\mu_{\rm B}^2} \vec{\mu}_i \cdot \Big(\sum_j J_{ij}\vec{\mu}_j \Big)- \mu_{iz}(H + H_{{\rm A}iz}),
\label{Eq:Ei}
\ee
where the factor of 1/2 in the first term reflects that the exchange energy is equally shared by $\vec{\mu}_i$ and its neighbors $\vec{\mu}_j$ and $J_{ij}$ is the exchange interaction between thermal-average spins ${\bf S}_i$ and ${\bf S}_j$, where $\vec{\mu} = -g\mu_{\rm B}{\bf S}$.  Our formulation of MFT does not use the concept of magnetic sublattices and is hence equally applicable to both collinear and planar noncollinear AFM structures in the magnetically ordered state.  The inclusions of anisotropy and finite field in the MFT are extensions of our published susceptibility calculations\cite{Johnston2012b} for zero anisotropy and zero (i.e., infinitesimal) applied field.  Our MFT is expressed in terms of physically measurable quantitities instead of in terms of the exchange interactions.  The anisotropy field is oriented along the easy ($z$) axis with $z$-component
\bse
\label{Eqs:HA0HA1}
\bea
H_{{\rm A}iz} &=& H_{\rm A0}\cos\theta_i,\ \ \bar{\mu} \equiv \frac{\mu}{\mu_{\rm sat}}, \ \ \mu_{\rm sat} = gS\mu_{\rm B},\label{Eq:HAiz}\\*
&&\hspace{0.3in} H_{\rm A0} = \frac{3H_{\rm A1}\bar{\mu}}{S+1}\label{Eq:HA0}
\eea
\ese
where $H_{\rm A0}\geq 0$ is the amplitude of the anisotropy field, $H_{\rm A1}\geq 0 $ is a subsidiary anisotropy field, $\theta_i$ is the angle between $\vec{\mu}_i$ and the $+z$ axis, $\mu$ is the magnitude of the ordered (and/or field-induced) moment at a given $H$ and $T$ and $\mu_{\rm sat}$ is the saturation moment at $T=0$.  For $H=0$, in the AFM phase we write
\be
\bar{\mu}(H=0,T)\equiv \bar{\mu}_0(T)\qquad ({\rm AFM\ phase}).
\ee
From the definition, $H_{{\rm A}iz}$ is oriented in the direction of the projection of $\vec{\mu}_i$ onto the $z$-axis and hence is generally different for different spins.  Since the anisotropy field acts in the same direction and therefore increases the effective exchange field that tends to align each ordered moment in the AFM structure at $H=0$, it is expected and found to increase the AFM ordering temperature at $H=0$, as discussed below for the collinear case.

For $H=H_{\rm A1} = 0$, for collinear or planar noncollinear AFMs one has\cite{Johnston2012b}
\bse
\label{Eqs:TNThetapf}
\bea
T_{\rm N} &=& -\frac{S(S+1)}{3k_{\rm B}} \sum_j J_{ij} \cos\phi_{ji},\label{eq:TN}\\*
\theta_{\rm p} &=& -\frac{S(S+1)}{3k_{\rm B}} \sum_j J_{ij},\label{eq:thetap}\\*
f &\equiv&\frac{\theta_{\rm p}}{T_{\rm N}} = \frac{\sum_j J_{ij}} {\sum_j J_{ij} \cos\phi_{ji}},\label{eq:f2}
\eea
\ese
where $\phi_{ji}$ is the angle between $\vec{\mu}_j$ and $\vec{\mu}_i$ in the AFM-ordered state with $H=0$.  We henceforth reserve the symbols $T_{\rm N}$, $\theta_{\rm p}$ and $f$ with the above definitions for this case of $H=H_{\rm A1} = 0$, which therefore do not depend on either the applied or anisotropy field.  When $H_{\rm A1} > 0$, we append ``A'' to these symbols, yielding $T_{\rm NA}$, $\theta_{\rm pA}$ and $f_{\rm A}$, respectively, which are then the observed values in this case.  Thus we define
\be
f_{\rm A} = \frac{\theta_{\rm pA}}{T_{\rm NA}}, \qquad t_{\rm A} = \frac{T}{T_{\rm NA}}.\nonumber\\*
\ee

We also define the dimensionless reduced anisotropy fields $h_{\rm A0}$ and $h_{\rm A1}$ and temperature $T_{\rm A1}$ as
\bse
\bea
h_{\rm A0} &\equiv& \frac{g\mu_{\rm B}H_{\rm A0}}{k_{\rm B}T_{\rm N}} = \frac{3\bar{\mu}h_{\rm A1}}{S+1}, \quad h_{\rm A1} \equiv \frac{g\mu_{\rm B}H_{\rm A1}}{k_{\rm B}T_{\rm N}}, \label{Eq:hA0hA1}\\*
&&T_{\rm A1} \equiv  \frac{g\mu_{\rm B}H_{\rm A1}}{k_{\rm B}},\quad h_{\rm A1} = \frac{T_{\rm A1}}{T_{\rm N}},\label{Eq:TA1Def}
\eea
where $T_{\rm N}$ is defined by the exchange constants and AFM structure in Eq.~(\ref{eq:TN}).  Then one finds
\bea
T_{\rm NA} &=& T_{\rm N}(1+h_{\rm A1}) = T_{\rm N}+T_{\rm A1},\quad \frac{T_{\rm NA}}{T_{\rm N}} - 1 = h_{\rm A1},\nonumber\\*
\theta_{\rm pA} &=& \theta_{\rm p}+T_{\rm A1},\qquad T_{\rm NA} - \theta_{\rm pA} = T_{\rm N} - \theta_{\rm p},\label{Eq:ThetapAThetap}
\eea
\ese
where $\theta_{\rm p}$ is defined in terms of the exchange interactions in Eq.~(\ref{eq:thetap}).  Thus $h_{\rm A1}$ has the appealing physical interpretation that it is equal to the relative enhancement of the AFM ordering temperature in $H=0$ due to the uniaxial anisotropy.  From Eqs.~(\ref{eq:f2}) and~(\ref{Eq:ThetapAThetap}) one sees that $f_{\rm A}$, $f$ and $h_{\rm A1}$ are not independent variables.  In particular, one has
\be
f_{\rm A} = \frac{f + h_{\rm A1}}{1 + h_{\rm A1}},\quad f = f_{\rm A}-h_{\rm A1}(1-f_{\rm A}).
\label{Eq:fAfhA1}
\ee

The magnetization per unit volume of \cacoas\ crystals even at saturation is sufficiently small that there is no need to correct the data for demagnetization effects.

\subsubsection{Magnetic Susceptibilities}

The magnetic susceptibilities per spin parallel ($\chi_{\parallel}$) and perpendicular ($\chi_{\perp}$) to the easy axis of a collinear AFM in the presence of the anisotropy field for temperatures at and above the observed N\'eel temperature $T_{\rm NA}$ are\cite{Johnston2012b}
\bse
\bea
\chi_\parallel(T) &=& \frac{C_1}{T-\theta_{\rm pA}},\quad \chi_\parallel(T_{\rm NA}) = \frac{C_1}{T_{\rm NA}-\theta_{\rm pA}}\\*
\chi_\perp(T) &=& \frac{C_1}{T-\theta_{\rm p}} = \frac{C_1}{T + T_{\rm A1} -\theta_{\rm pA}},\label{Eq:ChiPerp}\\*
\chi_\perp(T_{\rm NA}) &=& \frac{C_1}{T_{\rm NA}-\theta_{\rm p}}= \frac{C_1}{T_{\rm NA} + T_{\rm A1} -\theta_{\rm pA}}
\eea
\ese
where we have used Eqs.~(\ref{Eq:ThetapAThetap}) and the Curie constant per spin $C_1$ is given by
\be
C_1 = \frac{g^2S(S+1)\mu_{\rm B}^2}{3k_{\rm B}}.
\label{Eq:C1}
\ee
Thus in the presence of the uniaxial anisotropy $\chi_\parallel(T) > \chi_\perp(T)$ for all $T\geq T_{\rm NA}$, as expected, where $T_{\rm A1}$ is the amount by which the Weiss temperature in the Curie-Weiss law for parallel fields is changed towards positive values due to the presence of the anisotropy field.  On the other hand, when {\bf H} is perpendicular to the easy axis, the Weiss temperature $\theta_{\rm p}$ in Eq.~(\ref{Eq:ChiPerp}) is the same as if the anisotropy field were not present.

For $T\leq T_{\rm NA}$, one obtains
\bse
\label{Eqs:ChiParPerp}
\begin{eqnarray}
\chi_{\parallel}(T) &=& \left[ \frac{1-f_{\rm A}}{\tau^*_{\rm A}(t_{\rm A})-f_{\rm A}} \right] \chi_\parallel(T_{\rm NA}),
\label{eq:Chi_parallel}\\*
\chi_{\perp}(T) &=& \chi_{\perp}(T_{\rm NA}) ,
\end{eqnarray}
where 
\begin{eqnarray}
\tau^*_{\rm A}(t_{\rm A}) &=& \frac{(S+1)t_{\rm A}}{3B^\prime_S(y_{\rm 0A})}, \quad B^\prime_S(y) \equiv \frac{dB_S(y)}{dy},\\*
 y_{\rm 0A} &=& \frac{3\bar{\mu}_{\rm 0A}}{(S+1)t_{\rm A}}, \quad \bar{\mu}_{\rm 0A} \equiv \frac{\mu_{\rm 0A}}{\mu_{\rm sat}},\nonumber\\*
\mu_{\rm sat} &=& gS\mu_{\rm B},\quad\bar{\mu}_{\rm 0A} = B_S(y_{\rm 0A}).\nonumber
\end{eqnarray}
\ese
The subscript ``A'' in the above quantities again refers to their values in the presence of the uniaxial anisotropy. Note that the $g$-factor does not appear explicitly in Eqs.~(\ref{Eqs:ChiParPerp}). Equations~(\ref{Eqs:ChiParPerp}) are identical in form to those given  for $H_{\rm A1} = 0$ in Ref.~\onlinecite{Johnston2012b} with appropriate symbol replacements $f\to f_{\rm A}$, etc.  One has $1\leq \tau_{\rm A}^\ast < \infty$, $\tau_{\rm A}^\ast(t_{\rm A}\to 0) \to \infty$ and $\tau_{\rm A}^*(t_{\rm A}\rightarrow 1) = 1$, so Eq.~(\ref{eq:Chi_parallel}) gives $\chi_\parallel(t_{\rm A}\rightarrow 0) = 0$ and $\chi_\parallel(t_{\rm A}\rightarrow 1) = \chi_\parallel(T_{\rm NA})$ as expected.  From the possible range of $\tau_{\rm A}^\ast$ and the fact that $\chi_\parallel(T\leq T_{\rm NA})$ must be finite and positive for an antiferromagnet, Eq.~(\ref{eq:Chi_parallel}) gives the limits of $f_{\rm A}$ as
\begin{equation}
-\infty < f_{\rm A} < 1.
\label{Eq:fAlimits}
\end{equation}

However, the experimental values of $f_{\rm A}$ for \cacoas\ in Table~\ref{tab:CW} are greater than unity, in conflict with Eq.~(\ref{Eq:fAlimits}).  This disagreement suggests that spin fluctuations and correlations not accounted for by MFT significantly suppress $T_{\rm NA}$ from its MFT value, or that an itinerant antiferromagnet picture is more appropriate to \cacoas\ than the local moment picture we are considering in this section.

\begin{figure}
\includegraphics[width=3.3in]{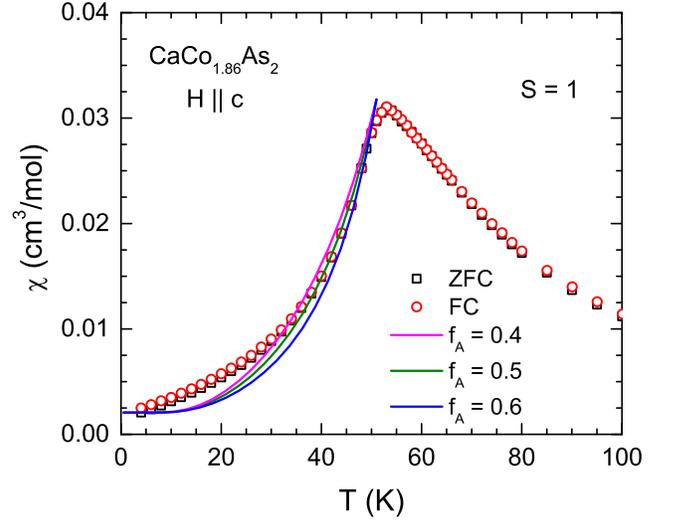}\vspace{0.1in}
\caption{(Color online) Zero-field-cooled (ZFC) and field-cooled (FC) magnetic susceptibility $\chi$ of a \cacoas\  single crystal as a function of temperature $T$ in the temperature range 4--100~K measured in a magnetic field $H= 0.01$~T applied along the $c$~axis ($\chi_c, H \parallel c$). The data below 4~K are excluded because they are affected by the superconductivity of adventitious Sn impurities on the surface of the crystal from the Sn flux used during crystal growth.  The solid curves are the mean-field theoretical predictions for $S=1$ and 0.4~$\leq f_{\rm A} \leq$~0.6.}
\label{fig:MT_CaCo2As2_A-type}
\end{figure}

The solid curves in Fig.~\ref{fig:MT_CaCo2As2_A-type} are the MFT predictions by Eq.~(\ref{eq:Chi_parallel}) for $\chi_{\parallel}(T)$ for a collinear A-type antiferromagnet with $S=1$ and $f_{\rm A} = 0.4$, 0.5 and 0.6. The curves are insensitive to the spin for $\frac{1}{2}\leq S \leq \frac{3}{2}$, so we took $S=1$ as a representative spin value in the plots.  The endpoint $\chi_\parallel(T = T_{\rm NA})$ of the fit is fixed by the experimental data and that at $T=0$ is fixed by the assumption of a collinear AFM\@. Although the fit is rather insensitive to the assumed value of $S$, only values of $f_{\rm A}$ in the range 0.4~$\leq f_{\rm A} \leq$~0.6 give reasonable agreement with the data. The value 
\be
f_{\rm A} = 0.5
\label{Eq:fAValue}
\ee
implies that in the absence of spin fluctuations and correlations not accounted for by MFT, one would have obtained $T_{\rm NA} = 129$~K using $\theta_{\rm pA}^c = 64.5$~K from Table~\ref{tab:CW}.

\subsubsection{Anisotropy Field from Critical Field Measurements}

A straightforward and accurate method of obtaining independent estimates of $f$ and $h_{\rm A1}$  is to model using MFT the observed critical fields if such data are available.  A critical field is defined here as the field at which the induced moment in the direction of the applied field becomes equal to the magnitude of the ordered moment at the same temperature.  MFT predicts that in the spin flop phase, i.e.\ for parallel fields above the spin flop field but below the spin flop critical field, the ordered moment is independent of field and the average induced $z$-component of the magnetic moment is proportional to field according to
\be
\mu_{\rm Flop} = \chi_{\rm Flop}H\qquad (H_{\rm Flop} \leq H \leq H_{\rm c\,Flop}),
\ee
where $\chi_{\rm Flop}$ is the magnetic susceptibility per spin which is independent of field in the spin-flop phase, and is understood to refer to the case where the applied field is parallel to the easy axis, which is the $c$~axis for \cacoas.  Similarly, when the field is applied perpendicular to the easy axis, the magnitude of the ordered moment is constant and the induced average perpendicular moment per spin $\mu_\perp$ is again proportional to {\bf H} from $H=0$ up to the perpendicular critical field $H_{\rm c\perp}$, according to
\be
\mu_\perp = \chi_\perp H,\qquad (0 \leq H \leq H_{\rm c\perp}).
\ee
The two susceptibilities are respectively given by
\bse
\bea
\chi_{\rm Flop} &=& \frac{C_1}{T_{\rm N}(1-f-h_{\rm A1})},\\*
\chi_{\perp} &=& \frac{C_1}{T_{\rm N}(1-f+h_{\rm A1})}.
\eea
\ese
Thus due to the anisotropy favoring parallel spin alignment, the parallel susceptibility in the spin flop phase is larger than the perpendicular susceptibility, in agreement with the data in Fig.~\ref{fig:MH_CaCo2As2_VSM}.

According to the above discussion, the respective critical field is defined as the field at which the induced moment equals the magnitude of the respective ordered moment at the given temperature, given by
\bse
\bea
H_{\rm c\,Flop} &=& \frac{\mu_{\rm Flop}}{\chi_{\rm Flop}},\\*
H_{\rm c\perp} &=& \frac{\mu_0}{\chi_\perp}.
\eea
\ese
In dimensionless reduced variables, one has
\bse
\label{Eqs:RedCritFields}
\bea
h_{\rm c\,Flop} &\equiv& \frac{g\mu_{\rm B}H_{\rm c\,Flop}}{k_{\rm B}T_{\rm N}} =\frac{3(1-f-h_{\rm A1})\bar{\mu}_{\rm Flop}}{S+1},\\*
h_{\rm c\perp} &\equiv& \frac{g\mu_{\rm B}H_{\rm c\perp}}{k_{\rm B}T_{\rm N}} = \frac{3(1-f+h_{\rm A1})\bar{\mu}_{\rm Flop}}{S+1},
\eea
\ese
The reduced ordered moment in the spin flop phase $\bar{\mu}_{\rm Flop}$ is not the same at a given temperature as the reduced ordered moment $\bar{\mu}_0$ in the AFM phase, except at $T=0$ at which $\bar{\mu}_{\rm Flop}= \bar{\mu}_0 = 1$.  Solving Eqs.~(\ref{Eqs:RedCritFields}) at $T=0$ for $f$ and $h_{\rm A1}$ gives
\bse
\label{fhA1fromhcs}
\bea
f &=& 1-\frac{S+1}{6}(h_{\rm c\perp} + h_{\rm c\,Flop}),\\* 
h_{\rm A1} &=& \frac{S+1}{6}(h_{\rm c\perp} - h_{\rm c\,Flop}).
\eea
\ese

From Fig.~\ref{fig:MH_CaCo2As2_VSM} the experimental critical field values for $T\to0$ are approximately
\be
H_{\rm c\,Flop} = 8.9~{\rm T}, \quad H_{\rm c\perp} = 11.2~{\rm T} \quad (T\to0).
\label{Eq:ExpHFlopHcValuesT0}
\ee
The saturation moment in Fig.~\ref{fig:MH_CaCo2As2_VSM} is about $\frac{1}{4}\,\mu_{\rm B}$/Co = $gS\mu_{\rm B}$/Co.  Therefore if we assume $S = 1/2$ one obtains
\be
g = 1/2,
\label{Eq:gexp}
\ee
which is very different from the value of $g\approx 1.6$ obtained in Eq.~(\ref{Eq:S12g}) from the Curie-Weiss fits to the anisotropic $\chi$ at high~$T$ and represents a serious inconsistency in the use of the local-moment model to fit the magnetic data for \cacoas.  However, continuing onward, using $T_{\rm N} = 52$~K the reduced experimental critical fields are
\be
h_{\rm c\,Flop} = 0.054,\qquad h_{\rm c\perp} = 0.071
\ee
Inserting these values into Eqs.~(\ref{fhA1fromhcs}) gives
\be
f = 0.969,\qquad h_{\rm A1}=0.0043.
\label{Eq:fhA1Exp}
\ee
The value of $f$ indicates a nearly FM system, qualitatively consistent with the FM intralayer alignment of the Co spins in the A-type AFM structure.  However, then Eq.~(\ref{Eq:fAfhA1}) predicts
\be
f_{\rm A} = 0.969,
\ee
instead of the value of 0.5 in Eq.~(\ref{Eq:fAValue}).  This is another of the inconsistencies that arise when using a local moment picture to model the magnetic properties of \cacoas.

\subsubsection{Anisotropy Field from Spin Flop Field Measurement}

For any collinear AFM containing identical crystallographically equivalent spins, the spin flop field $H_{\rm Flop}$ can be written in terms of the anisotropy field amplitude $H_{\rm A0}$ and the magnitude of the zero-field exchange field $H_{\rm exch0}$ seen by each spin as\cite{Johnston2012b}
\begin{equation}
H_{\rm Flop} = [(1-f)H_{\rm exch0} H_{\rm A0} - H_{\rm A0}^2]^{1/2}.
\label{Eq:HFlop}
\end{equation}
In standard treatments, $f$ is missing from this expression because the standard expression only applies to a bipartite AFM with only nearest-neighbor interactions that are the same.  In this case Eq.~(\ref{eq:f2}) yields $f=-1$.  Substituting $f=-1$ into Eq.~(\ref{Eq:HFlop}) gives the standard result.  Thus Eq.~(\ref{Eq:HFlop}) generalizes the standard result for an ideal bipartite AFM to real compounds which usually have $f \neq -1$.  Writing the fields in dimensionless form as in Eq.~(\ref{Eq:hA0hA1}), Eq.~(\ref{Eq:HFlop}) becomes
\begin{equation}
h_{\rm Flop} = [(1-f)h_{\rm exch0} h_{\rm A0} - h_{\rm A0}^2]^{1/2}.
\label{Eq:hFlop}
\end{equation}

From Eqs.~(\ref{Eq:Ei}) and~(\ref{eq:TN}), $H_{\rm exch0}$ for an antiferromagnet in $H=0$ can be written in standard and reduced forms, respectively, as
\bse
\bea
H_{\rm exch0} &=& \frac{T_{\rm N}}{C_1}\mu_0,\\*
h_{\rm exch0} &\equiv& \frac{g\mu_{\rm B}H_{\rm exch0}}{k_{\rm B}T_{\rm N}} = \frac{3\bar{\mu}_0}{S+1},
\eea
\ese
where $\mu_0$ is the magnitude of the ordered moment at a given $T$ in $H=0$, $\bar{\mu}_0\equiv \mu_0/\mu_{\rm sat}$, and the single-spin Curie constant $C_1$ is given in Eq.~(\ref{Eq:C1}).  Thus, at $T=0$ at which $\bar{\mu}_0=1$, one has simply
\bse
\label{Eqs:hexch0hA0T0}
\be
h_{\rm exch0}(T=0) = \frac{3}{S+1}.
\ee
From Eq.~(\ref{Eq:hA0hA1}), the reduced anisotropy fields $h_{\rm A0}$ and $h_{\rm A1}$ are related at $T=0$ by
\be
h_{\rm A0}(T=0) = \frac{3h_{\rm A1}}{S+1}.
\ee
\ese
Using Eqs.~(\ref{Eqs:hexch0hA0T0}), Eq.~(\ref{Eq:hFlop}) becomes
\begin{equation}
h_{\rm Flop} = \frac{3}{S+1}[(1-f) h_{\rm A1} - h_{\rm A1}^2]^{1/2} \quad (T=0).
\label{Eq:hFlopT0}
\end{equation}
Using the observed value $H_{\rm Flop}(T=0) = 3.5$~T from Fig.~\ref{fig:MH_CaCo2As2}(b), $T_{\rm N} = 52$~K and $g=1/2$ from Eq.~(\ref{Eq:gexp}) gives the dimensionless experimental spin flop field as $h_{\rm Flop}(T=0) = 0.022$.  Then using $f = 0.969$ from Eq.~(\ref{Eq:fhA1Exp}), Eq.~(\ref{Eq:hFlopT0}) gives the reduced anisotropy field $h_{\rm A1}(T=0)$ as
\be
h_{\rm A1}(T=0) = 0.0046.
\label{Eq:hA1Value}
\ee
This value is nearly the same as the value 0.0043 in Eq.~(\ref{Eq:fhA1Exp}) derived from the critical fields.

\subsubsection{Exchange Interactions}

For an A-type AFM structure with four FM (negative) nearest-neighbor interactions $J_1$ between moments in the $ab$-plane and two AFM (positive) nearest-neighbor interactions $J_c$ between a moment in one layer and the transition metal moments in the two adjacent layers of the ${\rm ThCr_2Si_2}$-type structure in Fig.~\ref{Fig:structure}, Eqs.~(\ref{Eqs:TNThetapf}) yield
\bse
\bea
T_{\rm N} &=& -\frac{2S(S+1)}{3k_{\rm B}} (2J_1 - J_c),\label{eq:J-TN}\\*
f &=& \frac{2J_1 + J_c}{2J_1 - J_c}.\label{eq:J-f}
\eea
\ese
Since $J_1<0$ and $J_c > 0$ one finds that $f < 1$.  Solving for $J_1$ and $J_c$ yields
\be
\bs
\frac{J_1}{k_{\rm B}} &= -\frac{3T_{\rm N}(1+f)}{8S(S+1)}, \\*
J_c &= -2J_1\frac{1-f}{1+f}.
\label{Eqs:J1JcFromExp}
\end{split}
\ee
Substituting the above experimental values $T_{\rm N}=52$~K, $S=1/2$ and $f = 0.969$ into Eqs.~(\ref{Eqs:J1JcFromExp}) yields
\be
\frac{J_1}{k_{\rm B}} = -52~{\rm K},\qquad \frac{J_c}{k_{\rm B}} = 1.7~{\rm K}.
\ee

Thus within the local-moment picture the exchange interactions are very anisotropic, with the in-plane FM interactions being about 30 times stronger than the interplane AFM interactions.  This anisotropy would be expected to be prominent in the spin-wave dispersion relations at low $T$\@.

\section{\label{Sec:CaCo2As2_HC} Heat Capacity}

\begin{figure}
\includegraphics[width=3in]{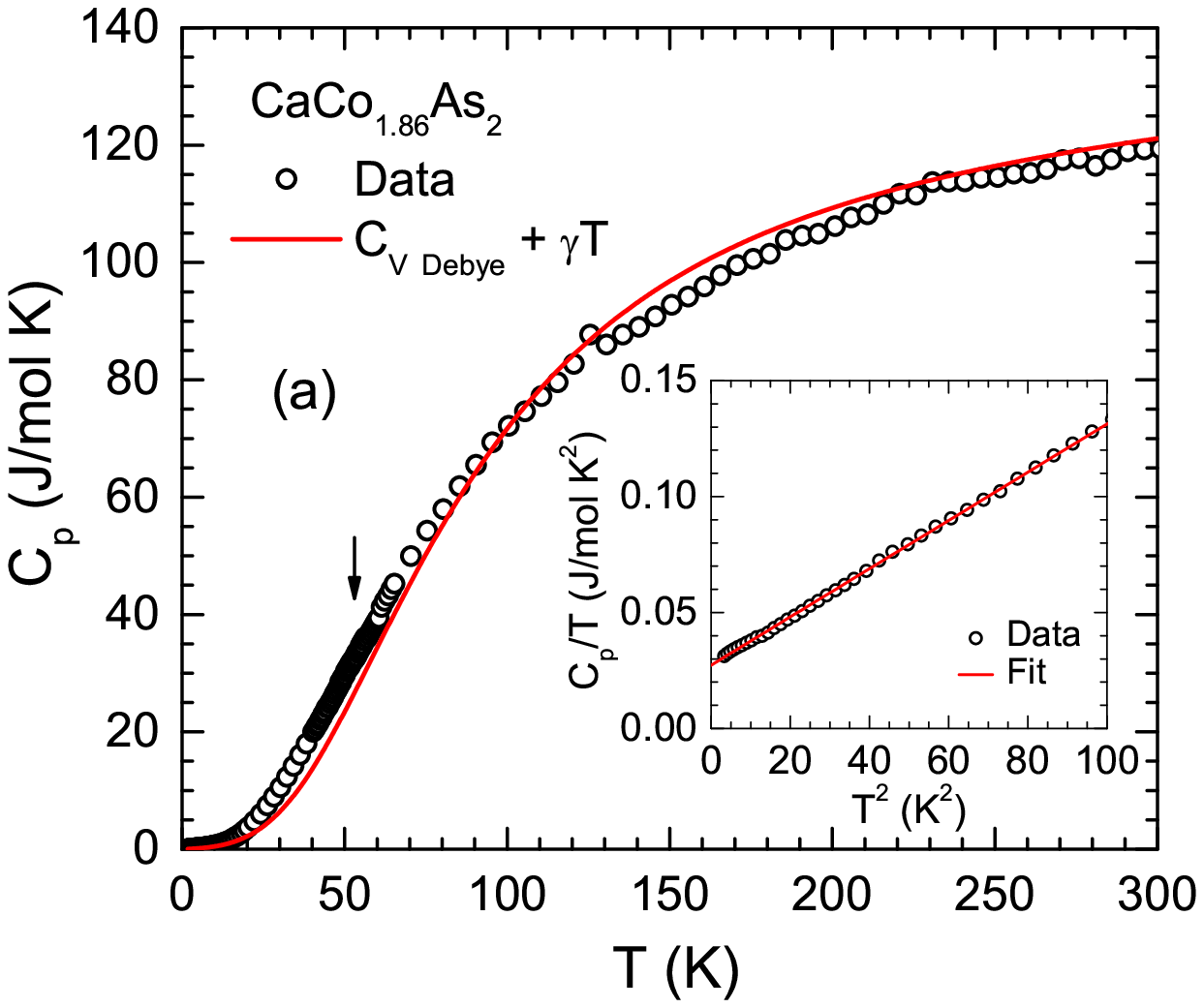}
\includegraphics[width=3in]{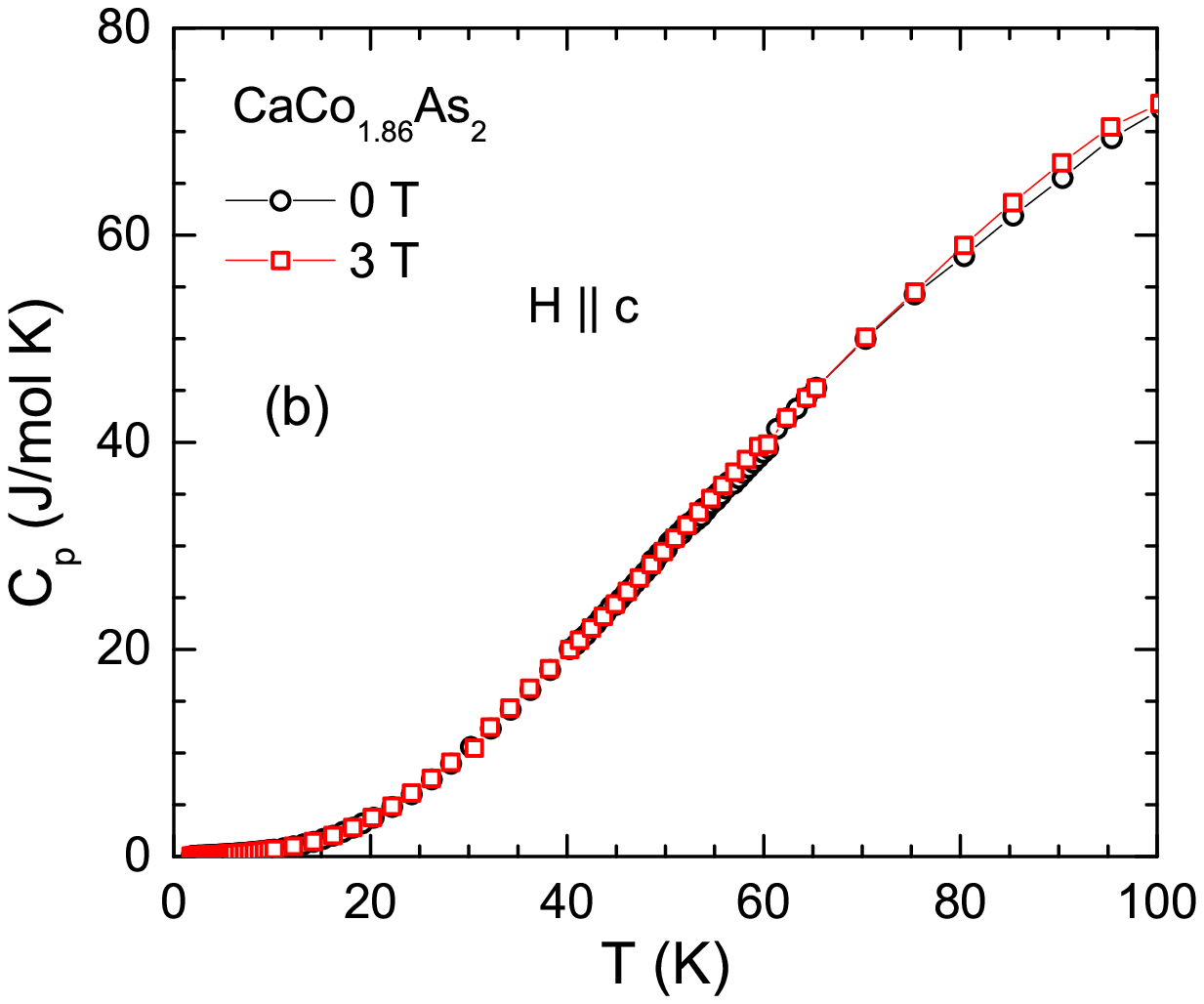}
\includegraphics[width=3in]{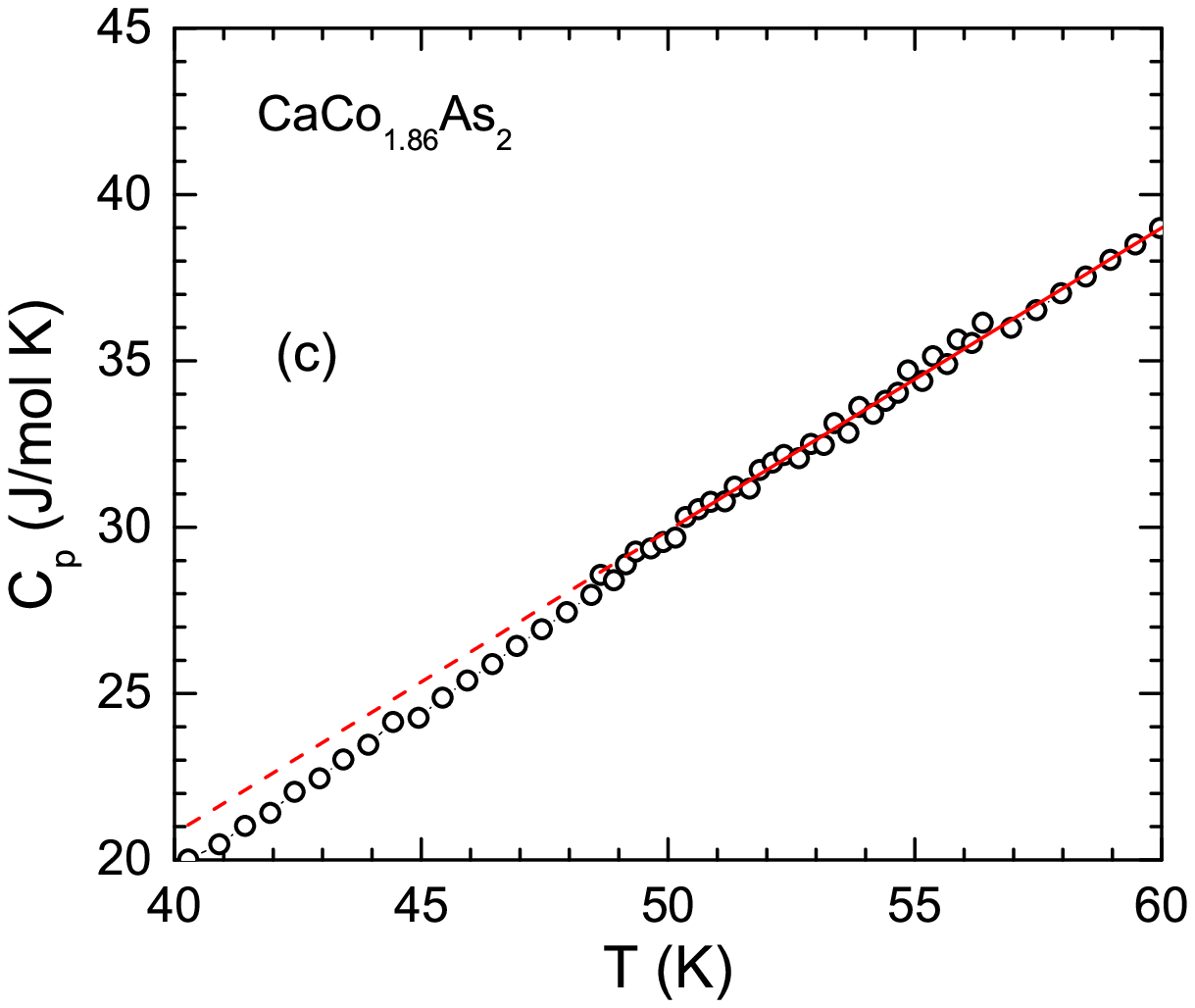}
\caption{(Color online) (a) Heat capacity $C_{\rm p}$ of a \cacoas\ single crystal as a function of temperature $T$ measured in zero magnetic field. The arrow marks $T = T_{\rm N}$, at which no obvious  anomaly is seen in $C_{\rm p}$. The solid curve is a fit by the sum of the contributions from the Debye lattice heat capacity $C_{\rm V\,Debye}(T)$ and electronic heat capacity $\gamma T$ according to Eq.~(\ref{eq:Debye_HC-fit}). Inset: $C_{\rm p}/T$ versus $T^2$ below 10~K\@.  The solid line represents the fit by Eq.~(\ref{Eq:CTFit}). (b)~$C_{\rm p}(T)$ of \cacoas\ for 1.8~K~$\leq T \leq$~100~K measured in $H = 0$ and 3.0~T ($H\parallel~c$).  (c)~Expanded plot of the data between 40 and 60~K, showing a possible small change in slope at about 50~K as indicated by the linear fit from 52 to 60~K (red solid line) and its extrapolation to lower $T$ (red dashed line).}
\label{fig:HC_CaCo2As2}
\end{figure}

The $C_{\rm p}(T)$ of a \cacoas\ crystal is shown in Fig.~\ref{fig:HC_CaCo2As2}.  
The heat capacity is found to attain a value $C_{\rm p} \approx 119$~J/mol\,K at room temperature which is close to the expected classical Dulong-Petit value $C_{\rm V} = 3nR = 14.58R$ = 121.2~J/mol\,K at constant volume, where $n=4.86$ is the number of atoms per f.u.\ and $R$ is the molar gas constant. \cite{Kittel2005, Gopal1966}  

The low-$T$ $C_{\rm p}(T)$ data below 10~K were analyzed using the conventional fit 
\be
\frac{C_{\rm p}(T)}{T} = \gamma + \beta T^2
\label{Eq:CTFit}
\ee
in the temperature range $1.8~{\rm K} \leq T\leq 10$~K, shown by straight red line in the inset of Fig.~\ref{fig:HC_CaCo2As2}(a).  The parameters obtained are
\be
\gamma=27(1)~{\rm \frac{mJ}{mol\,K^2}}, \quad \beta= 1.00(8)~{\rm \frac{mJ}{mol\,K^4}}. 
\label{Eq:gammabeta}
\ee
The Sommerfeld coefficient $\gamma$ is larger than those of the AFM-ordered 122-type iron-arsenide parent compounds ${\rm (Ca,Sr,Ba)Fe_2As_2}$ which range from 5.1 to 8.2~mJ/mol\,K$^2$ (Refs.~\onlinecite{Chen2008b,Ronning2008,Rotundu2010}) and indicates a large density of states at the Fermi energy ${\cal D}(E_{\rm F})$. 

In general, one can write
\be
\gamma = \gamma_0(1+\lambda_{\rm el-el})(1+\lambda_{\rm el-ph}),
\label{Eq:gamgam0}
\ee
where $\gamma_0$ is the bare band structure Sommerfeld coefficient, the factor $(1+\lambda_{\rm el-el})$ accounts for a possible many-body renormalization of the band structure (narrowing of the conduction bands) arising from an electron-electron interaction $\lambda_{\rm el-el}$ that is not included in the band structure calculations, $1+\lambda_{\rm el-ph}$ is the electron-phonon enhancement factor and $\lambda_{\rm el-ph}$ is the electron-phonon coupling constant.  Assuming that $\lambda_{\rm el-el} = \lambda_{\rm el-ph}=0$,  an upper limit of  ${\cal D}(E_{\rm F})$ is obtained from $\gamma$ using the relation\cite{Kittel2005} $\gamma_0 = (\pi^2 k_{\rm B}^2/3) {\cal D}(E_{\rm F})$, which gives 
\be
{\cal D}(E_{\rm F}) = 11.4(5)~{\rm \frac{states}{eV\,f.u.}\quad (both\ spin\ directions)}. 
\label{Eq:DEFValue}
\ee

The density of states calculated from the Sommerfeld coefficient of the low-$T$ heat capacity in Eq.~(\ref{Eq:DEFValue}) is significantly larger than the bare band structure density of states value is ${\cal D}(E_{\rm F}) = 2.95$~states/eV~f.u.\ for both spin directions  calculated for the observed A-type AFM structure in Eq.~(\ref{DEFAType}) below; the ratio between these two values is 3.86.  Then the product of the many-body band renormalization and electron-phonon enhancement factors in Eq.~(\ref{Eq:gamgam0}) gives
\be
(1+\lambda_{\rm el-el})(1+\lambda_{\rm el-ph}) = 3.86.
\label{Eq:EnhanceValue}
\ee
If one assumes $\lambda_{\rm el-el} = 0$, one would obtain an unrealistically large electron-phonon coupling constant $\lambda_{\rm el-ph} = 2.86$.  On the other hand, if one assumes $\lambda_{\rm el-el} = \lambda_{\rm el-ph}$, Eq.~(\ref{Eq:EnhanceValue}) yields $\lambda_{\rm el-el} = \lambda_{\rm el-ph} = 0.97$ with a more realistic value of $\lambda_{\rm el-ph}$.  The value of $\lambda_{\rm el-el}$ suggests a factor of two renormalization of the band structure (narrowing of the bands near the Fermi energy) due to many-body electron-electron interactions.

The Debye temperature is estimated from the value of $\beta$ in Eq.~(\ref{Eq:gammabeta}) using the relation \cite{Kittel2005} 
\be
\Theta_{\rm D} = \left(\frac{12 \pi^{4} n R}{5 \beta}\right)^{1/3},
\label{Eq:ThetaFromBeta}
\ee
which gives
\be
\Theta_{\rm D}= 212(1)~{\rm K},
\label{Eq:ThetaDValue}
\ee
where $n=4.86$ is again the number of atoms per formula unit.

The lattice heat capacity contribution to $C_{\rm p}(T)$ over the full $T$ range of the measurements was analyzed using
\begin{equation}
C_{\rm p}(T) = \gamma T + n C_{\rm{V\,Debye}}(T),
\label{eq:Debye_HC-fit}
\end{equation}
where $C_{\rm{V\,Debye}}(T)$ is the Debye lattice heat capacity at constant volume per mole of atoms due to acoustic phonons given by\cite{Gopal1966}
\begin{equation}
C_{\rm{V\,Debye}}(T) = 9 R \left( \frac{T}{\Theta_{\rm{D}}} \right)^3 {\int_0^{\Theta_{\rm{D}}/T} \frac{x^4 e^x}{(e^x-1)^2}\,dx}.
\label{eq:Debye_HC}
\end{equation}
In order to fit the $C_{\rm p}(T)$ data using the least squares method, the Debye lattice heat capacity integral was replaced by an analytic Pad\'e approximant function that greatly simplifies the fit.\cite{Ryan2012} While fitting the data we fixed $\gamma$ to the value $\gamma=27$~mJ/mol\,K$^2$ determined above in Eq.~(\ref{Eq:gammabeta}), and thus the $C_{\rm p}(T)$ data were fitted using only the one adjustable parameter $\Theta_{\rm D}$. The $C_{\rm p}(T)$ data were fitted by Eq.~(\ref{eq:Debye_HC-fit}) as shown by the solid red curve in Fig.~\ref{fig:HC_CaCo2As2}(a), yielding $\Theta_{\rm D} = 357(4)$~K\@. This value of $\Theta_{\rm D}$ is much larger than the value of 212~K obtained above in Eq.~(\ref{Eq:ThetaDValue}) from a fit to the low-$T$ $C_{\rm p}(T)$ data.  For a three-dimensional antiferromagnet the spin-wave contribution to the heat capacity has a $T^3$ dependence as does the lattice.  Therefore although a heat capacity anomaly is not obvious at $T_{\rm N}$, it seems likely that the $\beta$ value in Eq.~(\ref{Eq:gammabeta}) determined from the low-$T$ heat capacity contains a significant contribution from spin waves, hence yielding a $\beta$ value that is too large and a resultant $\Theta_{\rm D}$ value from Eq.~(\ref{Eq:ThetaFromBeta}) that is too small.

We carried out $C_{\rm p}(T)$ measurements in $H=3$~T $(H\parallel~c)$ as shown in Fig.~\ref{fig:HC_CaCo2As2}(b) where the data are compared with the data in $H=0$ from Fig.~\ref{fig:HC_CaCo2As2}(a).  One sees that there is no significant influence of the 3~T magnetic field on the data.

An expanded plot of the $C_{\rm p}(T)$ data between 40 and~60~K in Fig.~\ref{fig:HC_CaCo2As2}(a) is shown in Fig.~\ref{fig:HC_CaCo2As2}(c).  A small change in slope may occur near 50~K, i.e., near $T_{\rm N} = 52$~K, as suggested by the red solid and dashed lines in the figure.  The weakness of the anomaly in $C_{\rm p}(T)$ near $T_{\rm N}$ suggests that strong dynamic spin correlations may occur above $T_{\rm N}$ which would decrease the entropy associated with long-range ordering and hence the size of the heat capacity anomaly at $T_{\rm N}$.\cite{Mohn1989, Sun1991}

\section{\label{Sec:CaCo2As2Rho} Electrical Resistivity}

\begin{figure}
\includegraphics[width=3.3in]{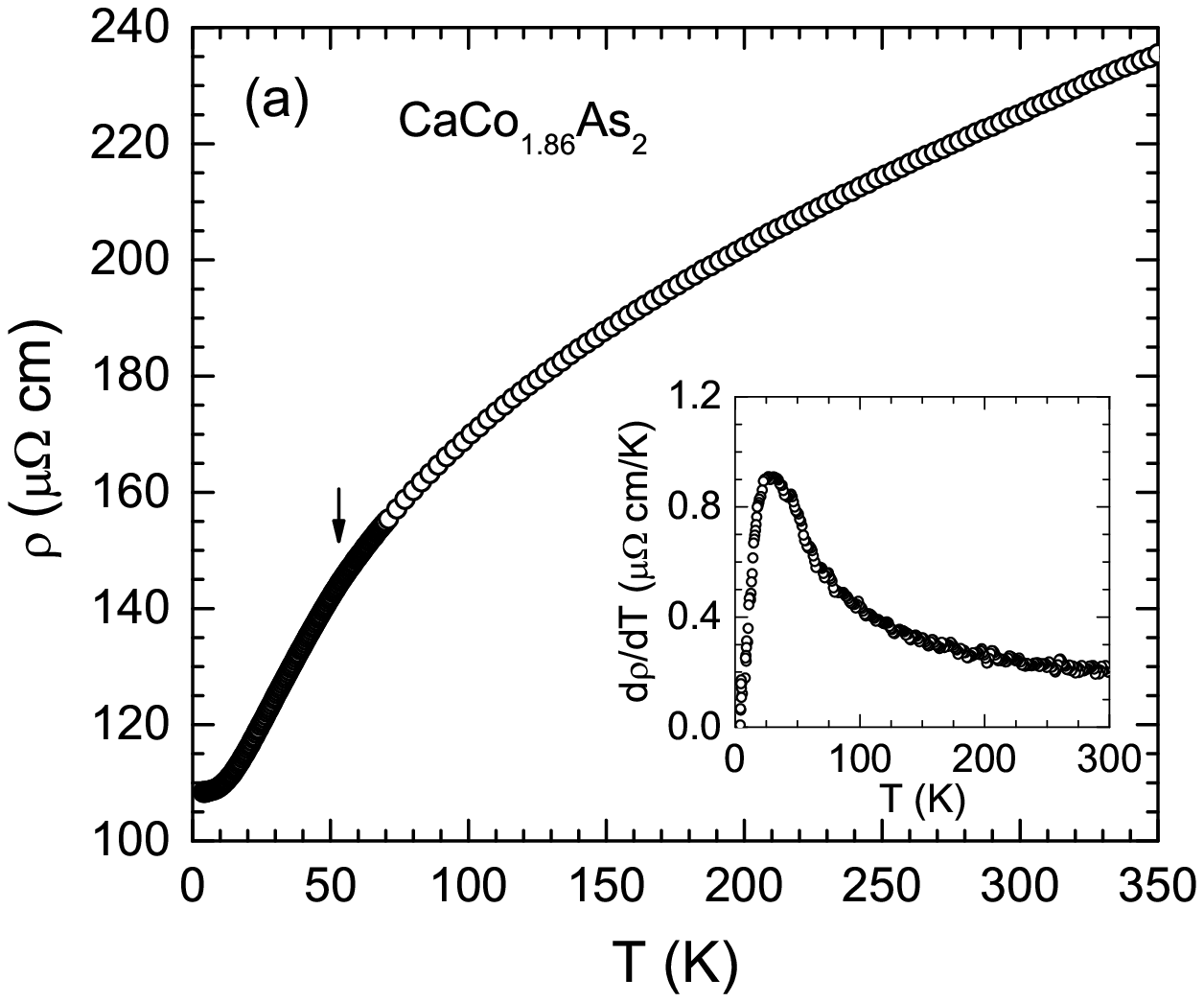}
\includegraphics[width=3.3in]{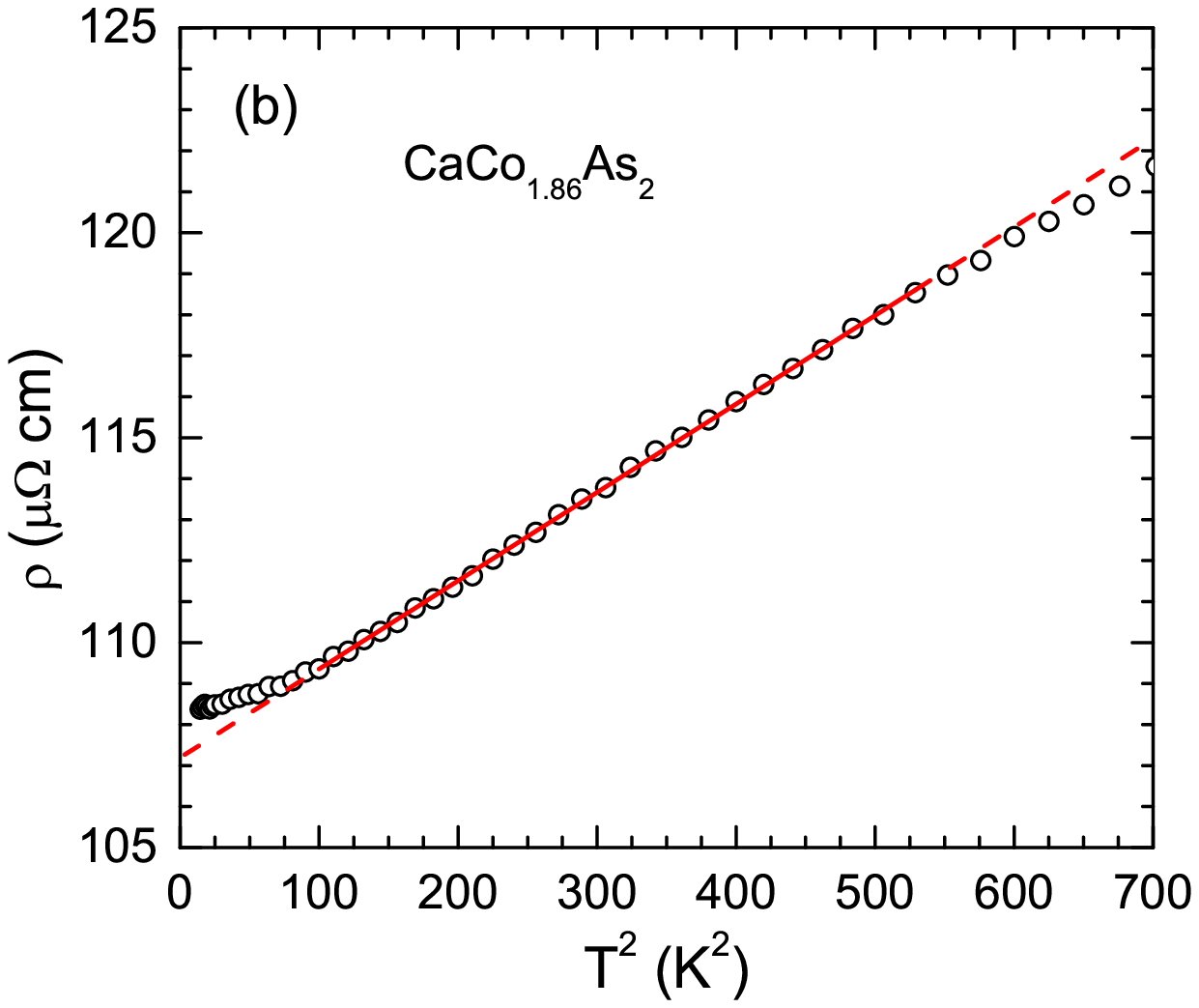}
\caption{(Color online) (a) In-plane electrical resistivity $\rho$ of a \cacoas\ single crystal as a function of temperature $T$ measured in zero magnetic field. The arrow marks $T = T_{\rm N} = 52$~K\@. Inset: Temperature derivative $d\rho/dT$ as a function of $T$\@. (b) $\rho$ vs $T^2$. The straight red line is a fit by $\rho = \rho_0' + A T^2$ over the temperature interval 10~K~$\leq T \leq$~23~K\@. The dashed lines are extrapolations of the fit.}
\label{fig:rho_CaCo2As2}
\end{figure}

The in-plane electrical resistivity $\rho$ of a ${\rm CaCo_2As_2}$ single crystal as a function of $T$ measured in zero magnetic field is shown in Fig.~\ref{fig:rho_CaCo2As2}. The $\rho(T)$ data exhibit metallic behavior with a residual resistivity $\rho_0 \approx 108~\mu \Omega\,{\rm cm}$ at $T= 4$~K and a residual resistivity ratio ${\rm RRR} \equiv \rho(300\,{\rm K})/\rho(4\,{\rm K}) \approx 2.2$. The data do not show any clear feature that might be associated with AFM ordering at $T_{\rm N}$. The absence of such a feature at $T_{\rm N}$ might be the result of very weak coupling between the Co moments and the conduction carriers. The derivative $d\rho(T)/dT$ is plotted versus $T$ in the inset of Fig.~\ref{fig:rho_CaCo2As2}(a), which also shows no clear feature at $T_{\rm N}$.

The $\rho(T)$ data exhibit a $T^2$ temperature dependence in the temperature range 10~K~$\leq T \leq$~23~K\@ as illustrated in the plot of $\rho$ vs $T^2$ in Fig.~\ref{fig:rho_CaCo2As2}(b). A fit of the $\rho(T)$ data by $\rho(T) = \rho_0' + A T^2$ in this $T$ range gives $\rho_0' = 107.1(1)~\mu \Omega\,{\rm cm}$ and $A = 2.18(3) \times 10^{-2}~\mu \Omega\,{\rm cm/K}^2$. The error bars do not take into account the systematic error of order 10\% arising from errors in the geometric factor.  The fit is shown in Fig.~\ref{fig:rho_CaCo2As2}(b).  However, since the $T^2$ dependence does not extend to $T\to0$, it may simply be a crossover within the overall temperature dependence of $\rho$ and hence not physically significant.

\section{\label{Sec:CaCo2As2ARPES} ARPES}

\begin{figure}
\includegraphics[width=\columnwidth]{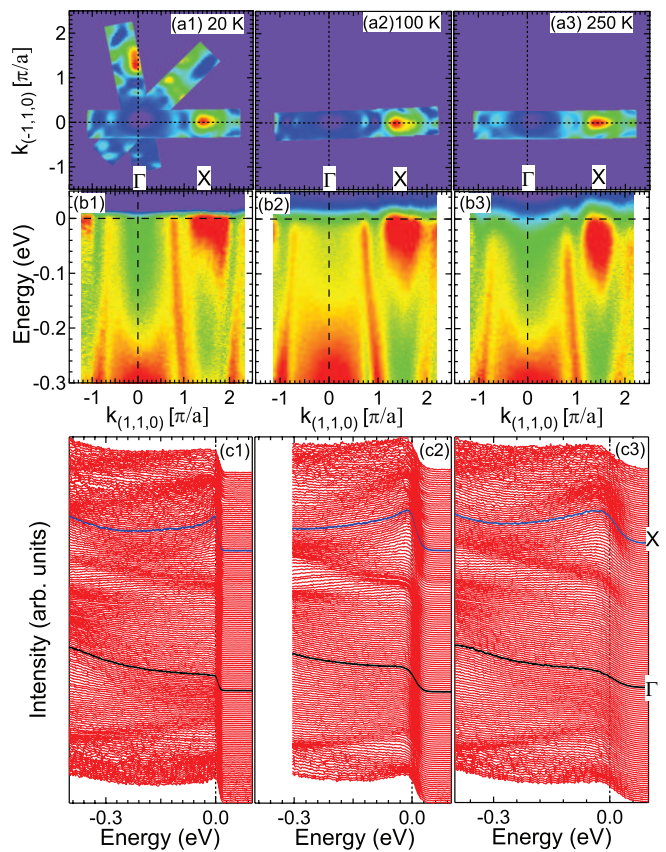}
\caption{(Color online) (a1, a2, a3) Fermi surface maps of \cacoas\ measured with a He~I photon source ($h \nu=21.2$~eV) at 20, 100 and 250~K sample temperatures, respectively. The band dispersion data along the [1,1,0] direction around $k_x=k_y=0$ and the corresponding electron distribution curves are shown in panels (b1, b2, b3) and (c1, c2, c3), respectively.}
\label{fig:ARPES}
\end{figure}

Figures~\ref{fig:ARPES}(a1--a3) show high-resolution ARPES intensity maps (Fermi surfaces of the \cacoas\ sample),  measured with 21.2~eV photon energy and at $\approx 20$~K, 100~K and 250~K sample temperatures, respectively. Each map is obtained by integrating the photoemission intensity over an energy window of $\pm 10$~meV centered at $E_{\rm F}$ and plotted as a function of in-plane wave vector components ($k_x$ and $k_y$). The Fermi surface map in Fig.~\ref{fig:ARPES}(a1) is shown in high symmetry directions [1,1,0], [0,1,0], and $[\bar{1}$,1,0], which indicates that there is no photoelectron intensity at the center of the Brillouin zone ($\Gamma$-point).  However, a huge intensity pocket is observed at the corner of the Brillouin zone (X-point).  One can examine the character of the Fermi surface pockets by tracing the dispersions of the associated bands along high symmetry lines. We have plotted the band dispersion near $E_{\rm F}$ along the [1,1,0] direction across the center of the Brillouin zone (BZ) and the corresponding energy distribution curves (EDCs), shown in Figs.~\ref{fig:ARPES}(b1) and \ref{fig:ARPES}(c1), respectively. Figure~\ref{fig:ARPES}(b1), measured at 20~K, clearly suggests the presence of an electron-like band dispersion at the $\Gamma$-point which does not cross $E_{\rm F}$, and the pocket at the X-point is also electron-like but does cross $E_{\rm F}$ and forms two Fermi surfaces. We calculated the Fermi momenta for the outer and inner electron pockets at the X-point, which are found to be about $\pm 0.7~{\rm \AA}^{-1}$ and $\pm 0.3~{\rm \AA}^{-1}$, respectively. The bottom of the inner electron pocket at the X-point is estimated to be at an energy of $\approx -0.1$~eV.   The Fermi velocity Êof the band shown in Fig.~\ref{fig:ARPES}(b1) is estimated as 3.00(4)~eV\,\AA. These results are consistent with previous ARPES studies of the similar compound ${\rm BaCo_2As_2}$.\cite{DhakaPRB13, Xu2013}

Recently, we observed significant changes in the band structure of the BaFe$_2$As$_2$ family with sample temperature using ARPES.\cite{Dhaka2011,Dhaka2013} Also relevant here is the AFM transition in \cacoas\ at about 52~K observed in our $\chi(T)$ measurements.  Therefore, to see whether the band structure changes across this transition, we have performed additional $T$-dependent ARPES measurements. Figures~\ref{fig:ARPES}(a2) and \ref{fig:ARPES}(a3) show the FS topology, measured at 100~K and 250~K, respectively. The corresponding band dispersions and EDCs are shown in Figs.~\ref{fig:ARPES}(b2, b3) and \ref{fig:ARPES}(c2, c3), respectively. These measurements indicate that there is no significant difference in the FS topology measured below [Fig.~\ref{fig:ARPES}(a1)] and above [Figs.~\ref{fig:ARPES}(a2) and \ref{fig:ARPES}(a3)] the transition temperature of 52~K. However, if we look at the band dispersion plots [Figs.~\ref{fig:ARPES}(b2) and \ref{fig:ARPES}(b3)], it seems that the bands are moving slightly downwards with increasing temperature.

\section{\label{Sec:CaCo2As2DOS} Electronic Structure Calculations}

\begin{figure}
\includegraphics[width=3.3in]{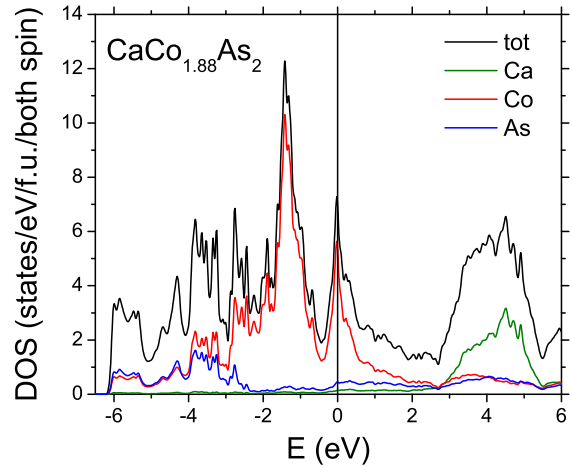}
\caption{(Color online) The density of states (DOS) for ${\rm Ca_8Co_{15}As_{16}}$ (${\rm CaCo_{1.875}As_2}$) as a function of energy $E$ relative to the Fermi energy $E_{\rm F}\equiv 0$ calculated using FPLAPW (black curve). The atom-decomposed DOS of the Ca, Co and As atoms are also shown.}
\label{fig:DOS}
\end{figure}

\begin{table}
\caption{\label{tab:ElecStruct} Results of electronic structure calculations for the paramagnetic (PM) state of ${\rm CaCo_{1.88}As_2}$, which is close to the measured composition ${\rm CaCo_{1.86}As_2}$, and for the idealized composition ${\rm CaCo_2As_2}$. The density of states at $E_{\rm F}$, ${\cal D}(E_{\rm F})$, is in units of states/(eV f.u.)\ for both spin directions.  Both the total and atom-decomposed ${\cal D}(E_{\rm F})$ values are shown where the values correspond to one Ca atom, two Co atoms and two As atoms.  Calculations of ${\cal D}(E_{\rm F})$ and total energy $E_{\rm tot}$ of the PM  and A-type, G-type and C-type antiferromagnetic (AFM) states relative to that of the ferromagnetic (FM) state from spin-polarized calculations are also listed.}
\begin{ruledtabular}
\begin{tabular}{lcc}
   & ${\rm CaCo_{1.88}As_2}$  & ${\rm CaCo_2As_2}$  \\
\hline
${\cal D}(E_{\rm F})$ (total, PM) & 7.29 & 8.50  \\
${\cal D}(E_{\rm F})$ (Ca, PM) & 0.15 & 0.16 \\
${\cal D}(E_{\rm F})$ (Co, PM) & 5.57 & 6.73 \\
${\cal D}(E_{\rm F})$ (As, PM) & 0.47 & 0.49 \\
\hline
${\cal D}(E_{\rm F})$ (total, A-type AFM) & 2.95 & 2.71\\
${\cal D}(E_{\rm F})$ (Ca, A-type AFM) & 0.12  &  0.11  \\
${\cal D}(E_{\rm F})$ (Co, A-type AFM) & 1.76  &  1.63  \\
${\cal D}(E_{\rm F})$ (As, A-type AFM) & 0.28  &  0.26  \\
\hline
${\cal D}(E_{\rm F})$ (total, FM)   & 3.22 & 2.73 \\
${\cal D}(E_{\rm F})$ (Ca, FM) & 0.12 & 0.12 \\
${\cal D}(E_{\rm F})$ (Co, FM) & 1.95 & 1.64 \\
${\cal D}(E_{\rm F})$ (As, FM) & 0.31 & 0.25 \\
\hline
$E_{\rm tot}$ (PM) (meV/f.u.) & 42.5    &  62.9 \\
$E_{\rm tot}$ (C-type AFM)  (meV/f.u.) &     &   63.2\\
$E_{\rm tot}$ (G-type AFM)  (meV/f.u.) &     &   62.4\\
$E_{\rm tot}$ (A-type AFM)  (meV/f.u.) &  2.64   &  3.2 \\
$E_{\rm tot}$ (FM)  (meV/f.u.) & $\equiv 0$  &  $\equiv 0$ \\

\end{tabular}
\end{ruledtabular}
\end{table}

To calculate the nonmagnetic (paramagnetic, PM), ferromagnetic (FM) and the various AFM ground states and total energies of a composition close to the experimental composition ${\rm CaCo_{1.86}As_2}$, we used a $2\times2\times 2$ supercell containing sixteen formula units (f.u.) with two vacancies on the Co sites to obtain the composition ${\rm Ca_8Co_{15}As_{16}}$ (${\rm CaCo_{1.88}As_2}$), which is close to the experimental composition.

The PM electronic density of states (DOS) as a function of energy $E$ relative to $E_{\rm F}$ calculated for ${\rm CaCo_{1.88}As_2}$ using FPLAPW is shown in Fig.~\ref{fig:DOS}, together with the atom-decomposed curves for Ca, Co and As. A sharp peak is observed in the total DOS at $E_{\rm F}$ mostly due to the contribution from the Co $3d$ bands. The atom-decomposed contributions to the DOS at $E_{\rm F}$ are given in the first four rows of Table~\ref{tab:ElecStruct} with a total DOS of 7.29 states/(eV f.u.)\ for both spin directions. The partial DOSs do not add up to the total DOS because of the interstitial DOS which is also accounted for in the calculation of the total DOS\@.  Thus the total DOS is higher than the sum of the atom-decomposed partial contributions.  As seen in Table~\ref{tab:ElecStruct}, corresponding calculations for the stochiomentric composition ${\rm CaCo_2As_2}$ give similar results.

To explore the theoretical ground states of \cacoastwo, we calculated the total energies of the PM, FM and the A-type, G-type and C-type AFM states [see Ref.~\onlinecite{Johnston2010} and Fig.~1(a) in Ref.~\onlinecite{Hotta1999} for the definitions of the AFM structure types] with 264 $k$-points in the irreducible Brillouin zone.  The total energies relative to the FM state for the A-type, G-type and C-type AFM and PM states are listed in Table~\ref{tab:ElecStruct}. These calculations show that the FM and A-type AFM states are the most stable with very similar total energies, which suggests the possibility of magnetic instability between these two states.  On the basis of these results, we calculated the total energies of only the PM and A-type AFM states relative to the FM state for ${\rm CaCo_{1.88}As_2}$ and the results are also listed in Table~\ref{tab:ElecStruct}.  The values of the total energies for ${\rm CaCo_{1.88}As_2}$ are similar to the corresponding ones for ${\rm CaCo_2As_2}$ and again indicate a close competition between the FM and A-type AFM states for the ground state.  A similar competition between itinerant FM and A-type AFM spin density wave states occurs in the parent compounds of the FeAs-based high-$T_{\rm c}$ superconductors.\cite{Johnston2010, Singh2008, Mazin2008}  

\begin{figure}
\includegraphics[width=3.in]{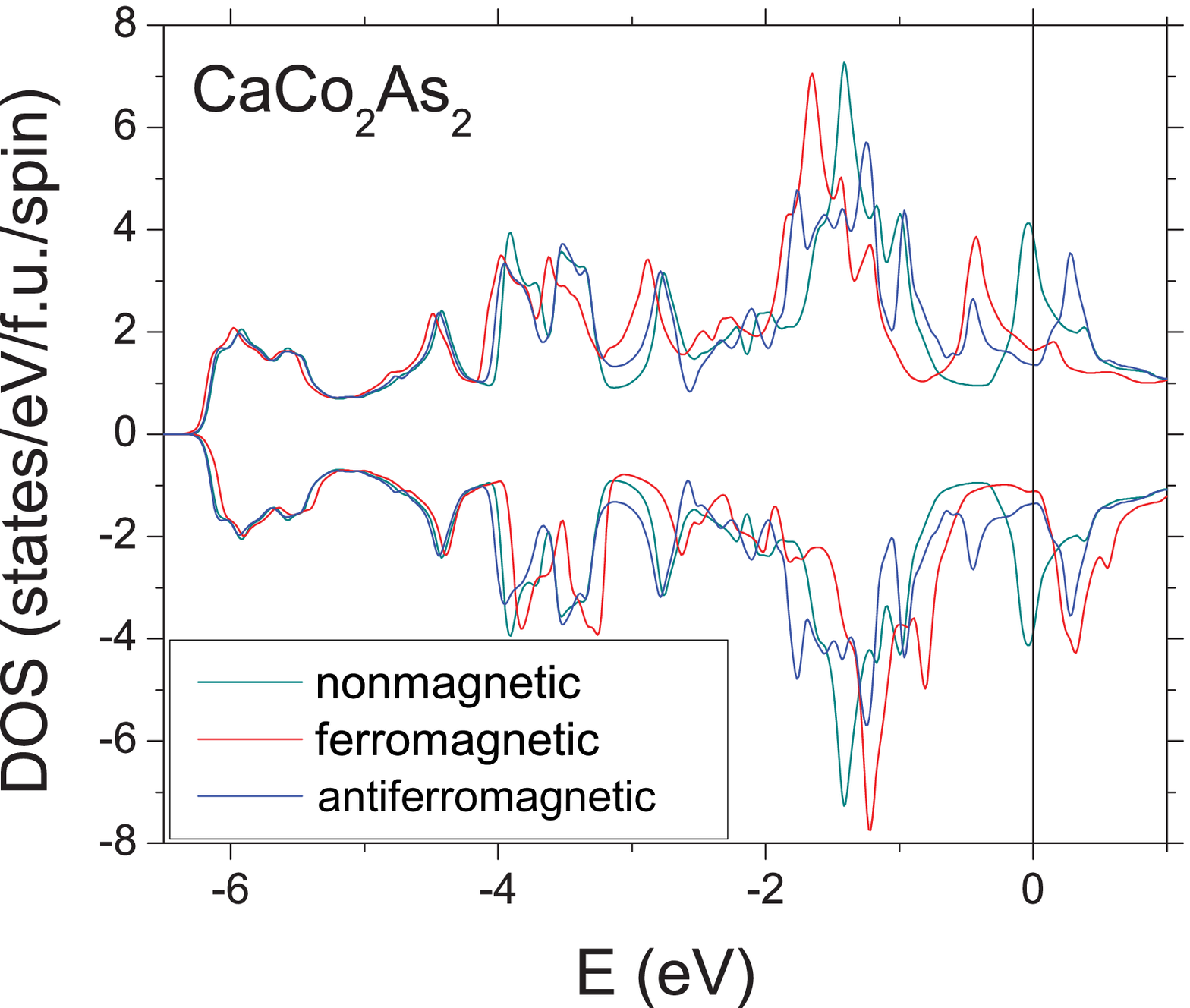}\vspace{0.1in}
\includegraphics[width=3.in]{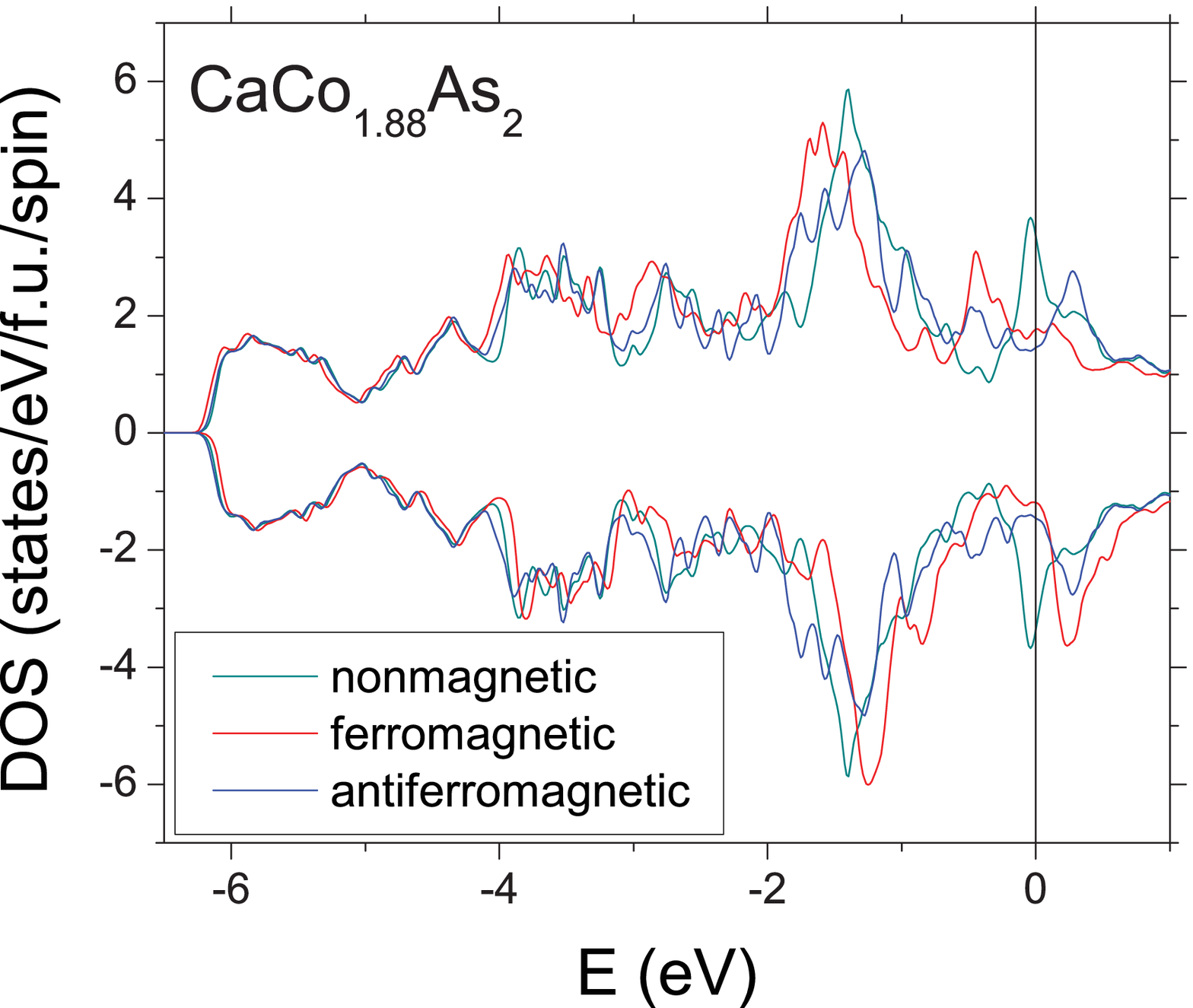}
\caption{(Color online) Total density of states (DOS) for ${\rm CaCo_2As_2}$ and ${\rm CaCo_{1.88}As_2}$ as a function of energy $E$ relative to the Fermi energy $E_{\rm F}\equiv 0$.  The calculations for each composition are for nonmagnetic, FM and A-type AFM states as indicated.  Positive DOS values are spin up and negative ones are spin down.  The nonmagnetic state shows a peak at $E_{\rm F}$ and this peak is split by both types of magnetic ordering for each composition.}
\label{fig:DOSMagOrd}
\end{figure}

A system that has a large DOS at $E_{\rm F}$ is often unstable and shows a tendency for a transition that splits the peak near $E_{\rm F}$, resulting in a decrease of the DOS at $E_{\rm F}$.  Spin-polarized calculations of the DOS versus energy of the PM, FM and A-type AFM states were carried out for both ${\rm CaCo_2As_2}$ and ${\rm CaCo_{1.88}As_2}$.  Figure~\ref{fig:DOSMagOrd} shows that both of these types of magnetic ordering indeed split the peak at $E_{\rm F}$ of the PM state and stabilize the system for each composition.  For the A-type AFM ordering observed in ${\rm CaCo_{1.86}As_2}$, the calculated total bare DOS at $E_{\rm F}$ for ${\rm CaCo_{1.88}As_2}$ decreases from
\bse
\label{Eqs:DEFPMAFM}
\be
{\cal D}(E_{\rm F}) = 7.29~{\rm \frac{states}{eV\,f.u.}\quad ( PM\ state})
\label{DEFPM}
\ee
for both spin directions in the PM state to
\be
{\cal D}(E_{\rm F}) = 2.95~{\rm \frac{states}{eV\,f.u.}\quad (A~type\ AFM})
\label{DEFAType}
\ee
\ese
for both spin directions in the A-type AFM state, which is only 40\% of the value calculated for the PM state.  The orbital decompositions of ${\cal D}(E_{\rm F})$ for the A-type AFM densities of states for ${\rm CaCo_2As_2}$ and ${\rm CaCo_{1.88}As_2}$ are listed in Table~\ref{tab:ElecStruct}.  Here again, the total DOS for each composition is larger than the sum of the orbital-decomposed DOSs due to the interstitial DOS\@.

\section{\label{Conclusions} Summary}

Structural studies on crushed \cacoas\ crystals demonstrated that the compound is in the collapsed-tetragonal phase of the body-centered tetragonal ${\rm ThCr_2Si_2}$-type structure.  Our refinement revealed a concentration of 6.0(5)\% vacancies on the Co sites, consistent with the value of 7(1)\% in Ref.~\onlinecite{Quirinale2013}.  The physical properties of \cacoas\ crystals were investigated by magnetic susceptibility, magnetization, specific heat, electrical resistivity and ARPES measurements and by electronic structure calculations.  The $\rho(T)$, $C_{\rm p}(T)$ and ARPES data and the band calculations indicate metallic behavior.  A large Sommerfield coefficient $\gamma = 27(1)$~mJ/mol\,K$^2$ and hence a large density of states at the Fermi energy is inferred from the $C_{\rm p}(T)$ data at low temperatures that is enhanced compared with the bare value predicted from the electronic structure calculations from the Co $3d$ bands.  Within an itinerant electron picture, a  large Stoner enhancement of the susceptibility by a factor of nine to two hundred is found, depending on the temperature.

A collinear A-type AFM ground state is inferred from the $\chi(T)$ and $M(H)$ data with $T_{\rm N} = 52$~K and with the easy axis being the $c$~axis, in agreement with the magnetic structure found from magnetic neutron diffraction measurements\cite{Quirinale2013} of \cacoas\ and inferred from previous $M(H,T)$ studies\cite{Cheng2012, Ying2012} of CaCo$_{2-x}$As$_2$.   A uniaxial anisotropy with an easy axis along the $c$~axis is deduced from the ordered-state $\chi(T)$ data, and $M(H)$ data exhibit a spin-flop transition ($H_{\rm Flop} =3.5$~T) for $H \parallel c$.  A spin-polarized total energy calculation showed that the ferromagnetic and A-type antiferromagnetic structures are in competition, where the experiments show that the latter wins out.  However, one must keep in mind that the A-type AFM structure consists of ferromagnetically-aligned layers of spins that are antiferromagnetically aligned between layers, and that the Curie-Weiss susceptibility at high~$T$ indicates that the dominant interactions are ferromagnetic.

An analysis of the $M(H,T)$ data in terms of a local moment model for the Co spins using molecular field theory did not yield self-consistent results, indicating that \cacoas\ is instead an itinerant electron antiferromagnet.  This assignment is consistent with the low ordered moment of $\approx 0.3~\mu_{\rm B}$/Co and the large exchange enhancement of the susceptibility.  No clear features are observed near $T_{\rm N}$ in our $\rho(T)$ and $C_{\rm p}(T)$ data.  The lack of an anomaly in $C_{\rm p}(T)$ at $T_{\rm N}$ suggests that strong short-range AFM order may occur above $T_{\rm N}$, thus removing most of the magnetic entropy on cooling to $T=T_{\rm N}$ and hence result in a small effect of long-range AFM ordering on the heat capacity at $T_{\rm N}$.

\acknowledgments

We thank A.~I. Goldman, A.~Kreyssig, R.~J. McQueeney and A.~Pandey for helpful discussions.  This research was supported by the U.S. Department of Energy, Office of Basic Energy Sciences, Division of Materials Sciences and Engineering.  Ames Laboratory is operated for the U.S. Department of Energy by Iowa State University under Contract No.~DE-AC02-07CH11358.

\end{document}